\documentclass[fleqn,usenatbib]{mnras}
\usepackage{newtxtext,newtxmath}
\usepackage[T1]{fontenc}
\usepackage{units}
\usepackage{lineno}
\usepackage{color}
\usepackage{epsfig} 
\usepackage{float}
\usepackage{bm}
\DeclareRobustCommand{\VAN}[3]{#2}
\let\VANthebibliography\thebibliography
\def\thebibliography{\DeclareRobustCommand{\VAN}[3]{##3}\VANthebibliography}
\usepackage{graphicx}
\usepackage{amsmath} 	 
 
\usepackage{amssymb}
\usepackage{amsfonts}
\usepackage{caption}
\usepackage{subfigure} 
\usepackage{multirow}
\usepackage[figuresright]{rotating}

\def\be{\begin{equation}}
\def\ee{\end{equation}}

\def\bes{\begin{align}}
\def\ees{\end{align}}

\newcommand{\fraction}[3]{\left(\frac{#1}{#2}\right)^{#3}}

\title[Coherent Curvature Radiation Spectrum by Fluctuating Bunches]{Coherent Curvature Radiation Spectrum by Dynamically Fluctuating Bunches in Magnetospheres} 

\author[Yang \& Zhang]{
Yuan-Pei Yang$^{1,2}$\thanks{E-mail: ypyang@ynu.edu.cn (YPY)}
and Bing Zhang$^{3,4}$\thanks{E-mail: bing.zhang@unlv.edu (BZ)}\\
$^{1}$South-Western Institute for Astronomy Research, Yunnan University, Kunming, Yunnan 650504, People's Republic of China\\
$^{2}$Purple Mountain Observatory, Chinese Academy of Sciences, Nanjing, Jiangsu 210023, People's Republic of China\\
$^{3}$Nevada Center for Astrophysics, University of Nevada, Las Vegas, NV 89154, USA\\
$^{4}$Department of Physics and Astronomy, University of Nevada, Las Vegas, NV 89154, USA\\
}

\date{Accepted XXX. Received YYY; in original form ZZZ}

\pubyear{2023}

\begin{document}
\label{firstpage}
\pagerange{\pageref{firstpage}--\pageref{lastpage}}
\maketitle

\begin{abstract}
Coherent curvature radiation by charged bunches has been discussed as the radiation mechanism for radio pulsars and fast radio bursts. Important issues for this radiation mechanism include how the bunches form and disperse in the magnetosphere of a pulsar or magnetar. More likely, bunches form and disperse continuously and it remains unclear what the spectral features are for these fluctuating bunches. In this work, we consider that the bunches in a magnetosphere have a formation rate of $\lambda_B$, a lifetime of $\tau_B$, and a typical Lorentz factor of $\gamma$, and analyze the spectral features of coherent curvature radiation by these fluctuating bunches. We find that the emission spectrum by a single fluctuating bunch is suppressed by a factor of $\sim(\lambda_B\tau_B)^2$ compared with that of a single persistent bunch, and there is a quasi-white noise in a wider band in the frequency domain. The high-frequency cutoff of the spectrum is at $\sim\max(\omega_{\rm peak},2\gamma^2/\tau_B)$, where $\omega_{\rm peak}$ is the peak frequency of curvature radiation. If the observed spectrum is not white-noise-like, the condition of $2\gamma^2\lambda_B\gtrsim \min(\omega_{\rm peak},2\gamma^2/\tau_B)$ would be required. Besides, the radiation by multiple fluctuating bunches along a field line is the incoherent summation of the radiation by single bunches if the bunch separation is longer than the wavelength. Conversely, a coherent summation should be involved. We also discuss the effects of bunch structures and the mechanism of bunch formation and dispersion. 
\end{abstract} 


\section{Introduction} 

The brightness temperatures of both radio pulsars and fast radio bursts (FRBs) are extremely high and are much greater than any plausible thermal temperature of the emitting electrons \citep[e.g.,][]{Melrose17,Petroff19,Cordes19,Zhang20,Zhang22b,Xiao21,Lyubarsky21,Bailes22}. This suggests that the radiation mechanism of radio pulsars and FRBs must be coherent.
For incoherent waves with random phases, the amplitude square of their superposition is approximately the sum of the amplitude squares of each wave. Thus, the observed emission power 
would be the simple summation of the emission power of individual charged particles, as proposed in most astrophysical scenarios. 
For coherent waves with certain phase differences, the amplitude square of their superposition could be significantly enhanced or reduced due to the wave coherent superposition process. In particular, ``coherently enhanced'' waves usually require that the phase differences of superposing waves must be much less the half wavelength of the waves.
In the literature about radiation mechanism, ``coherent'' is mainly defined as ``coherently enhanced''. 

Some coherent emission mechanisms have been invoked to interpret the emissions of radio pulsars and FRBs \citep[e.g.,][]{Melrose17,Zhang22b}: coherent radiation by charged bunches (i.e., antenna mechanism), maser by hydrodynamic instabilities or kinetic instabilities, etc. In this paper, we mainly focus on coherent curvature radiation by charged bunches that have been proposed as one of the popular ideas to explain the emission of pulsars \citep{Sturrock71,Ginzburg75,Ruderman75,Buschauer76,Benford77,Asseo98,Melikidze00,Gil04,Basu22,Rahaman22} and FRBs \citep{Katz14,Katz18,Kumar17,Lu18,Yang18,Kumar20,Lu20,Yang20,Cooper21,Wang22c,Wang22b,Tong22,Liu22,Qu23}.
Due to the two-stream instabilities in the magnetosphere of a neutron star, charged bunches might form and radiate electromagnetic radiation coherently \citep{Ruderman71,Benford77,Cheng77,Usov87,Rahaman20,Kumar22b}. 
However, as pointed out by some authors \citep[e.g.,][]{Melrose17,Lyubarsky21}, these models involving charged bunches have some important issues: (1) The charged bunches might be short-lived; (2) The radiation might be strongly suppressed by the magnetosphere plasma. For the latter issue, \citet{Gil04} and \citet{Lyubarsky21} pointed out that electromagnetic waves with frequencies below the plasma frequency could propagate in the highly magnetized plasma in the magnetosphere, but the radiation power would be significantly suppressed. However, \citet{Qu23} recently found that the plasma suppression effect could be ignored in the case of FRBs because of the existence of a parallel electric field in the FRB emission region, as is required to power the bright FRB emission.
The former issue leads to a more fundamental question: Is it necessary that the charged bunches have to be long-lived in order to explain the observed features of radio pulsars and FRBs? In other words, how do the formation and dispersion of bunches affect the observed radiation? 

In this work, we will analyze the spectral features of the coherent curvature radiation by dynamically fluctuating bunches in the magnetosphere of a neutron star. We consider that the bunches in the magnetosphere form with an average rate of $\lambda_B$ and have an average lifetime of $\tau_B$, and discuss how $\lambda_B$ and $\tau_B$ affect the radiation spectral feature. The paper is organized as follows. In Section \ref{section_brightness}, we discuss the brightness temperature of curvature radiation in a magnetosphere using a more physical treatment. In Section \ref{section_bunch}, we analyze the spectral features of coherent curvature radiation by fluctuating bunches, including the features by a single persistent bunch with different structures (Section \ref{subsection_persistent}), a single fluctuating bunch (Section \ref{subsection_fluctuating}), and multiple fluctuating bunches (Section \ref{subsection_multiple}). In Section \ref{section_formation}, we discuss the formation and dispersion mechanisms of bunches in the magnetosphere and calculate $\lambda_B$ and $\tau_B$. The results are summarized and discussed in Section \ref{section_discussion}. The convention $Q_x=Q/10^x$ is adopted in cgs units unless otherwise specified.

\section{Brightness temperature: the difference between observational treatment and physical treatment}\label{section_brightness}

Before discussing the spectral features of coherent curvature radiation by charged bunches, we point out the differences between the observational treatment and the physical treatment of brightness temperature. 
The observational brightness temperature is usually defined by 
\begin{align}
T_B = \frac{1}{2\pi k_B}\left(\frac{d}{\nu\Delta t}\right)^2 F_\nu=10^{35}~{\rm K}~d_{\rm Gpc}^2\nu_9^{-2}\Delta t_{-3}^{-2}F_{\nu,{\rm Jy}},\label{brightness}
\end{align}
where $k_B$ is the Boltzmann constant, $d$ is the distance between source and observer\footnote{Here we do not specify whether $d$ is luminosity distance or angular diameter distance for an order of magnitude estimate. A more precise treatment involves a correction factor with a certain power of $(1+z)$, see \cite{Luo23} and \cite{Zhang22b} for details.}, $\Delta t$ is the duration of a transient, $\nu$ is the frequency of the electromagnetic wave, $d_{\rm Gpc}=d/1~{\rm Gpc}$ and $F_{\nu,{\rm Jy}}=F_\nu/1~{\rm Jy}$. Since all quantities in Eq.(\ref{brightness}) are observable, this brightness temperature could be directly estimated from the observations \citep[e.g.,][]{Xiao22,Luo23,Zhu-Ge23}. However, this observationally-defined brightness temperature is based on a toy model that the transverse size of the emission region is approximately $c \Delta t$, which does not apply to broader astrophysical scenarios, which we discuss more below.

We first briefly summarize the meaning of the physical brightness temperature.
For a photon field within a volume element $dV$ and a solid angle element $d\Omega$, the number of states per volume per frequency interval per solid angle is given by \citep[e.g.,][]{Rybicki86}
\be
dN_s=\frac{2dVd^3k}{(2\pi)^3}=\frac{2\nu^2}{c^3}dVd\nu d\Omega,\label{state}
\ee
where $d^3k=k^2dkd\Omega$ is the three-dimensional wave vector element, and the factor 2 accounts for the fact that photons have two independent polarizations per wave vector $\vec{k}$. 
As bosons, the average number of photons in a state, i.e. the occupation number, is 
\be
\mathcal{N}=\frac{1}{\exp(h\nu/kT)-1},
\ee
where $T$ is the temperature of the photon field.
Therefore, the directional energy density $u_\nu(\Omega)$ is given by
\be
u_\nu(\Omega) dVd\nu d\Omega=\frac{2\nu^2}{c^3}h\nu \mathcal{N} dVd\nu d\Omega.
\ee
For radio observations, the brightness temperature $T_B$ is usually defined under the Rayleigh-Jeans limit, $I_\nu=2\nu^2kT_B/c^2$. Using $u_\nu(\Omega)=I_\nu/c$, one finally has
\be
kT_B=\mathcal{N}h\nu.
\ee
Thus, \emph{the brightness temperature directly reflects the photon energy per state.} 
The physical implications of an extremely large brightness temperature include the following: (1) For incoherent radiation, there is a maximum brightness temperature a gas can reach without undergoing Compton catastrophe \citep[e.g.,][]{Longair11,Melrose17}, and for a synchrotron self-absorbed radio source, this maximum brightness temperature is $\sim 10^{12}~{\rm K}$; 
(2) Induced Compton scattering would be significant at a high brightness temperature because photons satisfy Bose-Einstein statistics \citep[e.g.,][]{Lyubarsky16,Lu18}; (3) The interaction between electromagnetic waves and particles/plasmas in the ambient medium would be non-linear due to the particles' relativistic motion in strong electromagnetic waves \citep[e.g.,][]{Yang20b}.

Physically, the flux-intensity relation is generally given by
$F_\nu=\pi I_\nu\left(l_e/d\right)^2$,
where $l_e$ is \emph{the emission region transverse scale perpendicular to the line of sight}.
The brightness temperature can be therefore written as
\be
T_B = \frac{c^2}{2\pi k_B\nu^2}\left(\frac{d}{l_e}\right)^2 F_\nu.
\ee
This equation can be reduced to the observationally defined brightness temperature Eq.(\ref{brightness}) when $l_e = c \Delta t$ is satisfied. For this equation to be satisfied, three conditions are needed:
(1) The emission region is non-relativistic; 
(2) Its longitudinal scale is the same order of magnitude as the transverse scale; 
(3) The radiation is isotropic at any point in the emission region.

In realistic physical conditions, the transverse scale $l_e$ is model dependent and could be different from $c \Delta t$. The physical brightness temperature could be very different within different physical scenarios given the same observations. A well-known effect that has been discussed in the literature is when the emitter has a relativistic bulk motion, e.g. within the scenarios invoking a relativistic outflow \citep[e.g.,][]{Lyubarsky14,Beloborodov17,Metzger17}.
The above-mentioned condition 1 and condition 2 would not be satisfied. 
The transverse scale of the emission region $l_e$ is about $l_e\simeq2\Gamma^2c\Delta t/\Gamma = 2 \Gamma c \Delta t$, where $\Gamma$ is the Lorentz factor of the outflow. Due to relativistic motion, the observable emission region of the outflow is within the angle of $1/\Gamma$ pointing to the observer.
Then the physical brightness temperature is
\begin{align}
T_B&=\frac{1}{8\pi k_B}\left(\frac{d}{\nu\Gamma\Delta t}\right)^2 F_\nu\nonumber\\
&=2.7\times10^{32}~{\rm K}~d_{\rm Gpc}^2\nu_9^{-2}\Delta t_{-3}^{-2}\Gamma_1^{-2}F_{\nu,{\rm Jy}},
\end{align}
which is much smaller than that given by the traditional formula in Eq.(\ref{brightness}).

Here we point out that the physical brightness temperature in magnetopheric models, including the curvature radiation scenario discussed in this paper, is also different from the traditional definition and should be properly redefined. 

We consider the case that the emission region is within a magnetosphere and charged particles are moving relativistically, as are envisaged in many theoretical models of radio pulsars and FRBs \citep[e.g.,][]{Sturrock71,Ruderman75,Kumar17,Yang18,Lu20}. 
The above-mentioned condition (2) and (3) would not be satisfied. 
We consider that the charged particles/bunches are relativistically moving with a Lorentz factor $\gamma$ along the curved field lines with a curvature radius $\rho$. For $\omega\lesssim\omega_c$, where $\omega_c\sim\gamma^3c/\rho$ is the typical frequency of curvature radiation, the radiation beaming angle at the angular frequency $\omega$ is approximately \citep{Jackson98}
\be
\theta_e(\omega)\sim\frac{1}{\gamma}\fraction{\omega_c}{\omega}{1/3},
\ee 
which involves the field line direction.
Thus, the transverse lengthscale $l_e$ of the emission region at the distance $r$ from the neutron star center should be estimated by
\begin{align}
l_e&\sim r\theta_e(\omega)\sim\frac{3}{4}\rho\theta\theta_e(\omega)\nonumber\\
&\sim\frac{3}{4}\theta\fraction{c\rho^2}{\omega}{1/3}\simeq2.7\times10^5~{\rm cm}~\rho_8^{2/3}\nu_9^{-1/3}\theta,
\end{align}
where $\rho\simeq 4r/3\theta$ is the curvature radius at the position $(r,\theta)$, and $\theta$ is the poloidal angle between the emission region and the magnetic axis. 
We can see that for the above typical parameters, the transverse lengthscale $l_e$ is much smaller than that estimated by $c\Delta t\sim3\times10^7~{\rm cm}~\Delta t_{-3}$.
As a result, a more physical brightness temperature for the curvature radiation can be estimated as 
\begin{align}
T_B&=\frac{32\pi}{9k_B}\fraction{d}{\theta}{2}\fraction{c}{\rho\omega}{4/3}F_\nu\nonumber\\
&=1.3\times10^{39}~{\rm K}~d_{\rm Gpc}^2\theta^{-2}\rho_8^{-4/3}\nu_9^{-4/3} F_{\nu,{\rm Jy}}.
\end{align}
One can see that for curvature radiation in a magnetosphere, the physical brightness temperature should depend on the emission region parameters $(\rho,\theta)$. For typical parameters, the brightness temperature is much larger than the value estimated by the traditional formula Eq.(\ref{brightness}). Meanwhile, it is worth noting
that such a brightness temperature is independent of the burst duration $\Delta t$. The reason is that the burst duration does not directly reflect the transverse size of the emission region.

\section{Coherent curvature radiation by fluctuating bunches}\label{section_bunch}

In the magnetosphere of a neutron star, charged bunches form and disperse continuously.
Charged bunches are thought to be formed by certain plasma instabilities in the magnetosphere, e.g., two-stream instability \citep{Ruderman75,Benford77,Cheng77,Usov87,Asseo98,Melikidze00,Rahaman20}, and be dispersed by electrostatic repulsion, velocity dispersion, etc.
In this section, we mainly calculate the spectrum of the coherent curvature radiation by fluctuating bunches and discuss how the spectral feature is affected by the bunch formation and dispersion.

\begin{figure}
    \centering
	\includegraphics[width = 1.0\linewidth, trim = 100 150 100 120, clip]{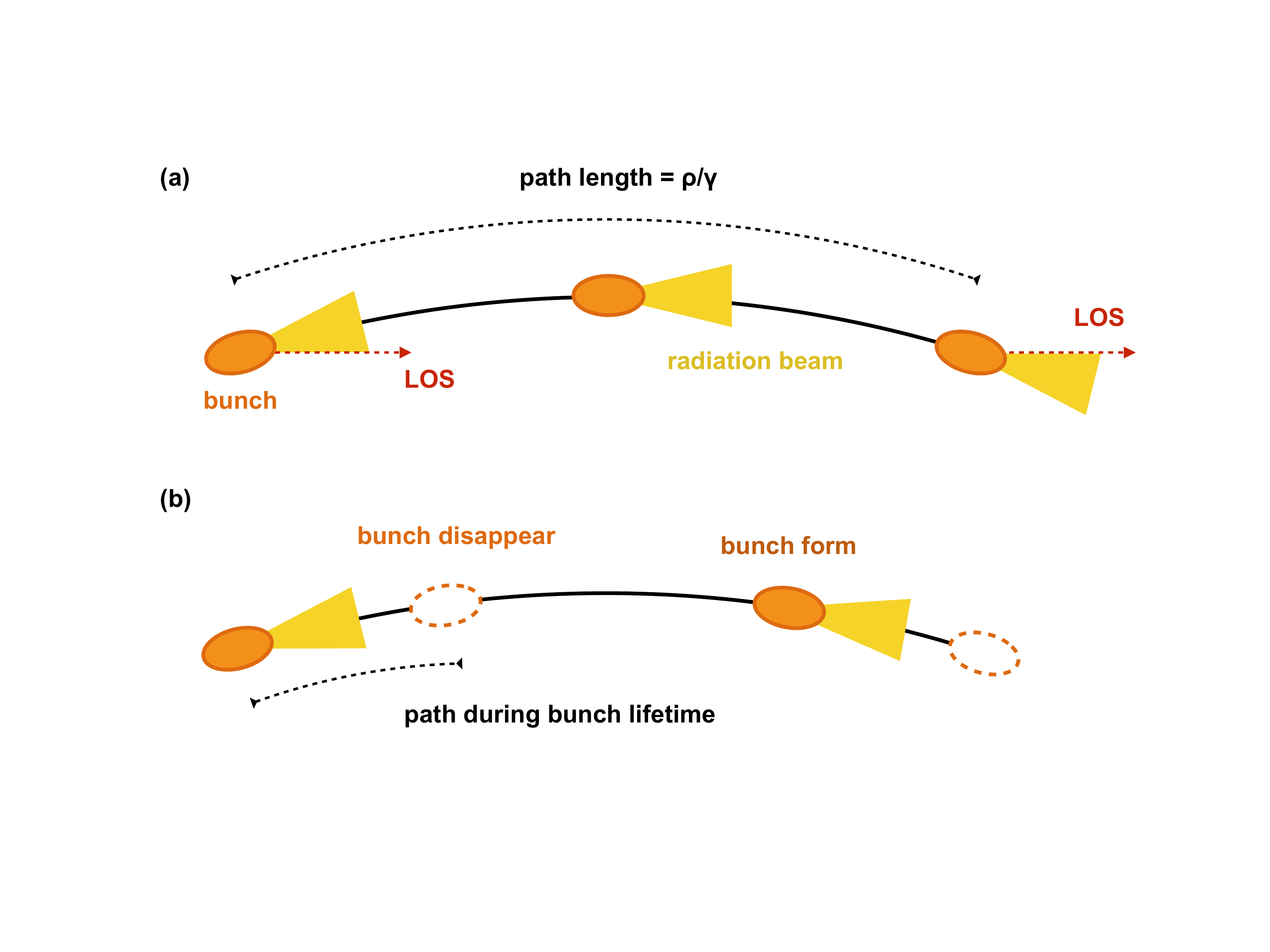}
    \caption{Schematic configurations of a bunch moving along a magnetic field line and emitting curvature radiation.
    The black line denotes the magnetic field line, the orange ellipse denotes the bunch, the dashed ellipse denotes the bunch disappearing due to dispersion, and the yellow region denotes the radiation by the bunch due to relativistic motion.
    Panel (a): the bunch is persistent. Due to the relativistic motion of the bunch, the length scale of the path along the line of sight (LOS) is $\rho/\gamma$, where $\rho$ is the curvature radius of the field line, and $\gamma$ is the bunch Lorentz factor.
    (b) the bunch is fluctuating due to rapid formation and dispersion when the building plasma moves along the field line.}\label{bunch_fluctuation} 
\end{figure}

\subsection{Radiation by a single persistent bunch with different structures}\label{subsection_persistent}
First, we briefly summarize the spectral properties of a single persistent bunch. We consider that the bunch has the velocity $v$ with Lorentz factor $\gamma=(1-v^2/c^2)^{-1}$ and moves along the magnetic field line with a curvature radius $\rho$, as shown in the panel (a) of Figure \ref{bunch_fluctuation}. 

If the bunch lifetime is long enough, $\tau_B>\rho/\gamma c$, which is comparable to the time of a persistent bunch sliding along a curved magnetic field line, the observer will see the radiation with the emission cone of angular width $\sim1/\gamma$ around the observer direction, and the typical angular frequency of the emission wave is
\be
\omega_c=\frac{1}{\tau_c}\sim\left[\frac{\rho}{\gamma c}\left(1-\frac{v}{c}\right)\right]^{-1}\simeq\frac{2\gamma^3c}{\rho}, \label{frequency1}
\ee
where $\tau_c$ is the typical pulse duration of the classical curvature radiation for a single point source, and the factor of $(1-v/c)$ is due to the propagation time-delay effect.

We consider that the classical curvature radiation is in the form of a finite pulse $E(t)$, and $E(t)$ vanishes sufficiently rapidly for $t\rightarrow\pm\infty$. For convenience, we define $A(t)\equiv(c/4\pi)^{1/2}[RE(t)]_{\rm ret}$, where $R$ is the distance between the observer and the bunch at the retarded time, and the bracket $[...]_{\rm ret}$ is evaluated at the retarded time.
The Fourier transform of $A(t)$ is defined as 
\begin{align}
&A(\omega)=\frac{1}{\sqrt{2\pi}}\int_{-\infty}^{\infty}A(t)e^{i\omega t}dt,\label{Aomega}\\
&A(t)=\frac{1}{\sqrt{2\pi}}\int_{-\infty}^{\infty}A(\omega)e^{-i\omega t}d\omega.\label{At}
\end{align}
Here we adopt the above definitions of Fourier transforms as the same as those in \citet{Jackson98}, and the corresponding properties of Fourier transform will be adopted accordingly in the following discussion. 
The directional emission spectrum (defined as \emph{the radiation energy per unit solid angle per unit angular frequency}) is \citep{Jackson98}
\be
P_A(\omega)\equiv\frac{dW}{d\Omega d\omega}
=2|A(\omega)|^2
=\frac{c}{4\pi^2}\left|\int_{-\infty}^{\infty}[RE(t)]_{\rm ret}e^{i\omega t}dt\right|^2.\label{PAomega}
\ee

\begin{figure}
    \centering
	\includegraphics[width = 1.0\linewidth, trim = 40 50 20 50, clip]{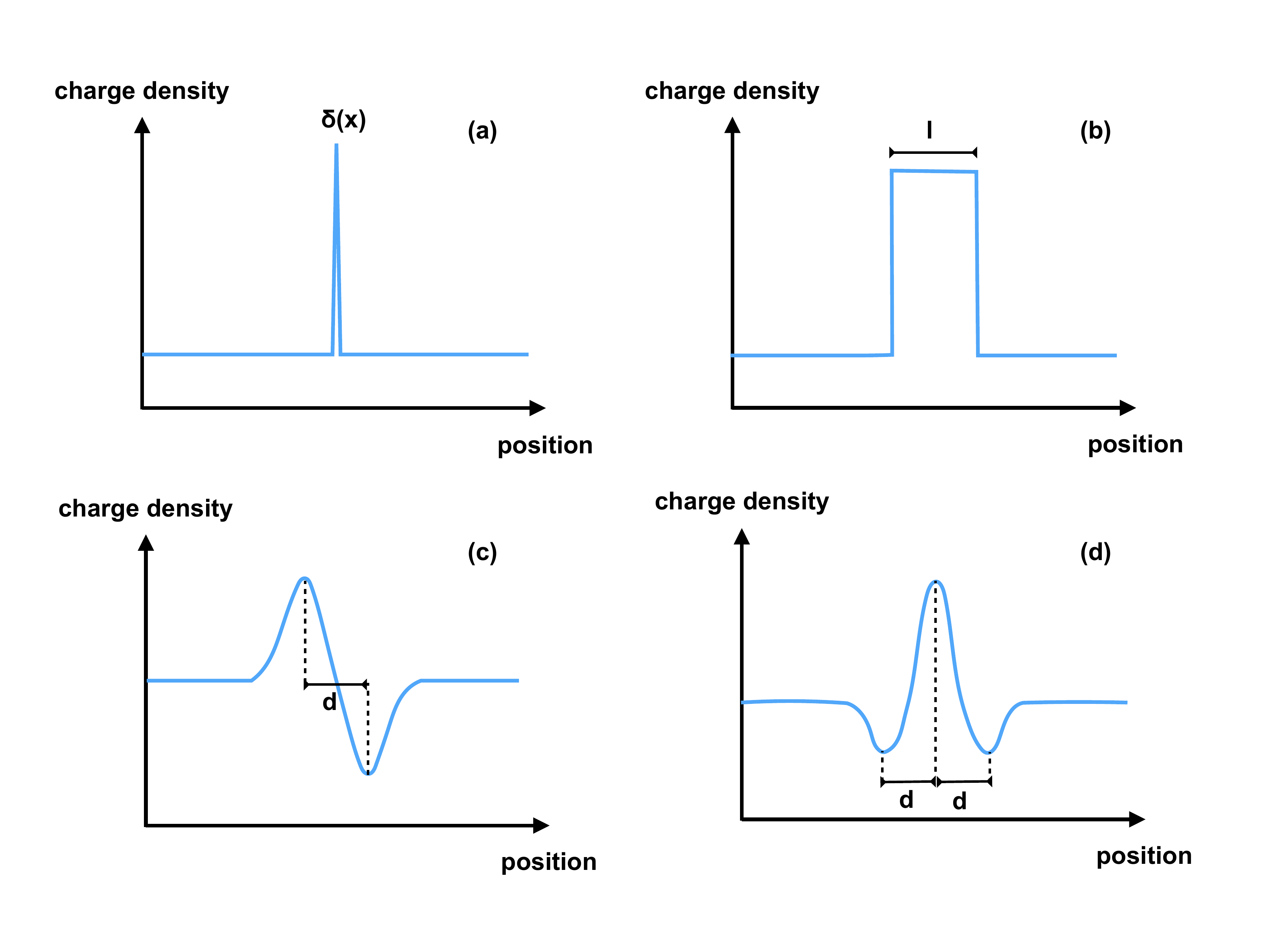}
    \caption{Charge density distributions for different bunch structures. Panel (a): A single point-source bunch, with the charge density described by a delta function $\delta(x)$; Panel (b): A one-dimensional bunch with lengthscale $l$, with a uniform charge density; Panel (c): A bunch-cavity pair formed in a plasma background, with the separation between the bunch and the cavity being $d$. Panel (d): A bunch-cavity system satisfying the structures of a soliton, with the separations between the bunch and the cavities being $d$.}\label{bunch_shape} 
\end{figure}

\begin{figure}
    \centering
	\includegraphics[width = 1.0\linewidth, trim = 0 0 0 0, clip]{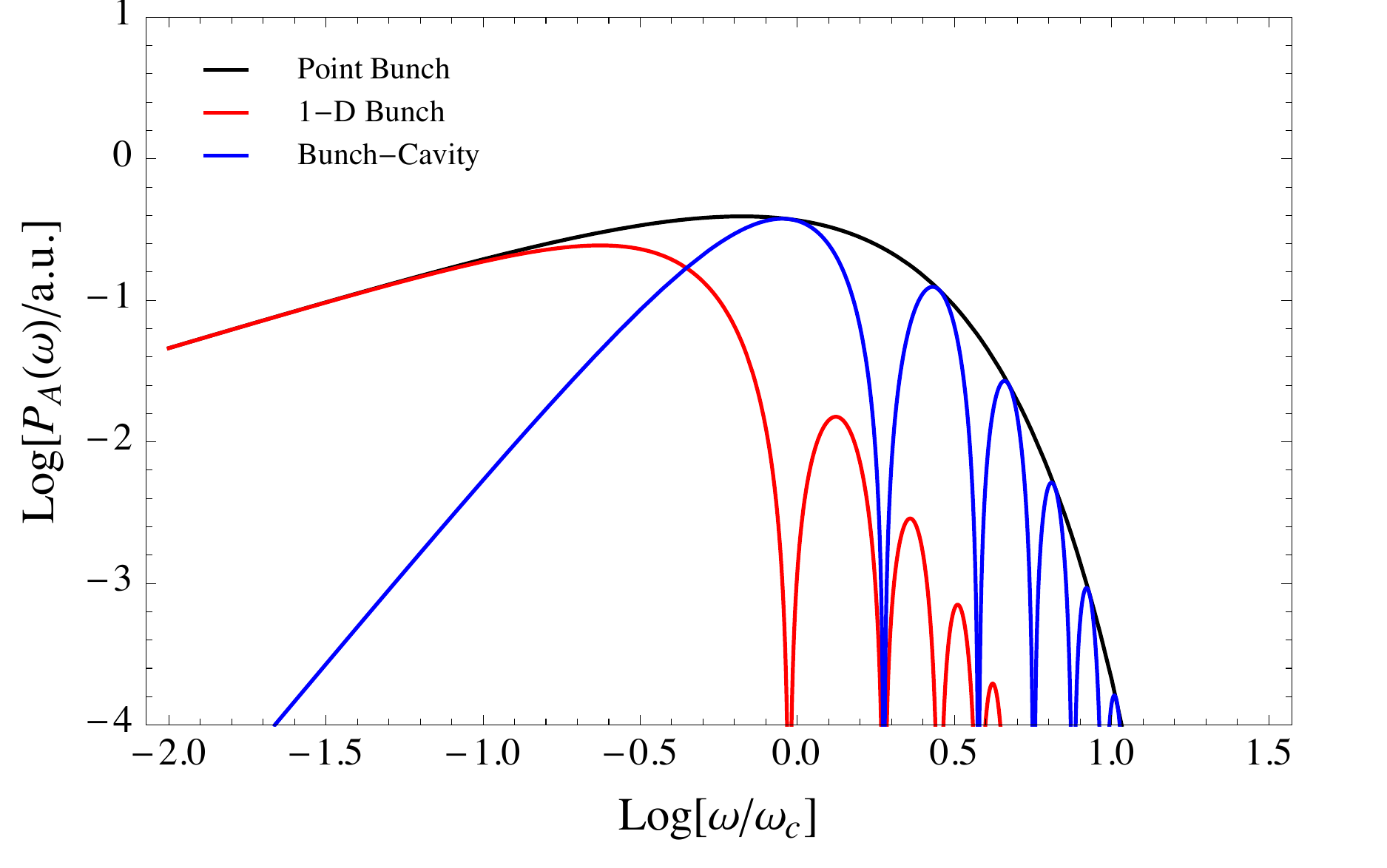}
    \caption{The emission spectrum of a single persistent bunch. The black, red, and blue curves correspond to the emission spectrum of a point-source bunch (panel (a) in Figure \ref{bunch_shape}), a one-dimensional bunch with $\omega_l=0.3\omega_c$ (panel (b) in Figure \ref{bunch_shape}), a bunch-cavity system (a bunch-cavity pair (panel (c) in Figure \ref{bunch_shape}) or a soliton (panel (d) in Figure \ref{bunch_shape})) with $\omega_d=0.3\omega_c$, respectively. The unit of the emission spectrum is arbitrary. For easy comparison with the spectral shapes of different scenarios, the emission spectrum of the bunch-cavity system is suppressed by an arbitrary factor in this figure.}\label{PA_spectra} 
\end{figure}

For the curvature radiation by a single persistent bunch, the properties of $P_A(\omega)$ mainly depend on the spatial structure of the charged bunch \citep{Yang18,Yang20}. We briefly summarize the following three scenarios with the different structures as shown in Figure \ref{bunch_shape}:

1. A point-source bunch (see the panel (a) of Figure \ref{bunch_shape}): If the bunch length is much smaller than a half wavelength, the point-source approximation is reasonable. The emission spectrum of the curvature radiation is \citep{Jackson98,Yang18}
\be
P_A(\omega)\propto\omega^{2/3}e^{-\omega/\omega_c}.\label{CRspec}
\ee
In particular, the spectral index of ${2/3}$ is due to the angular spectrum involved in the curvature radiation, which is different from the scenario of synchrotron radiation that is usually described by total spectrum with a spectral index of $1/3$, except the case with a narrow pitch-angle distribution \citep{Yang18b}. The emission spectrum is shown by the black curve in Figure \ref{PA_spectra}.

2. A one-dimensional bunch with a finite length $l$ (see the panel (b) of Figure \ref{bunch_shape}): The corresponding emission spectrum of curvature radiation is \citep{Yang18}
\be
P_A(\omega)\propto{\rm sinc}^2\left(\frac{\omega}{\omega_l}\right)\omega^{2/3}e^{-\omega/\omega_c}~~~\text{with }\omega_l\simeq2c/l,\label{1Dbunch}
\ee
where ${\rm sinc}(x)\equiv\sin x/x$ that is involved due to the coherent summation of the phase factor of the charged particles at different regions on the one-dimensional bunch (also see the similar derivation of Eq.(\ref{phasefactor})), $\omega_l$ is the angular frequency corresponding to the half length of a bunch. Here the charge distribution of the bunch is assumed to be uniform. If $\omega\gg\omega_l$, one has ${\rm sinc}^2(\omega/\omega_l)\sim \omega^{-2}$ due to the rapid oscillation of the term $\sin^2(\omega/\omega_l)$ with a unit amplitude in the ${\rm sinc}$ function, leading to a softer spectrum compared with that of a point source. The emission spectrum is shown by the red curve in Figure \ref{PA_spectra}.
In particular, when $\omega_l<\omega_c$,
the peak radiation specific power would be suppressed by a factor of $\sim(\omega_l/\omega_c)^{2/3}$, leading to the total radiation energy suppressed by a factor 
\be
\eta_{l}\simeq\frac{\omega_l P_A(\omega_l)}{\omega_c P_A(\omega_c)}\simeq\fraction{\omega_l}{\omega_c}{5/3}\simeq0.9l_1^{-5/3}\nu_{c,9}^{-5/3},
\ee
compared with that of a point source given by Eq.(\ref{CRspec}). This formula can be used to estimate how the bunch length suppresses the total radiation power. 

3. A bunch-cavity pair or similar system formed by plasma background fluctuation (see panel (c) and panel (d) of Figure \ref{bunch_shape}). 
First, we consider that a charged bunch forms in the plasma background and has a charge density larger than the background, then a corresponding cavity with a charge density smaller than the background would form near the bunch, as shown in panel (c) of Figure \ref{bunch_shape}. 
For simplicity, we treat the bunch-cavity pair as a two-point source with a separation $d$. 
Thus, the charge density distribution of the bunch-cavity pair system could be described by $\rho_{\rm bc}(x)=q\delta(x)-q\delta(x-d)+\rho_0$, where $x$ denotes the pair position, $\pm q$ in the first two terms correspond to the charges of the bunch and the cavity, respectively, and $\rho_0$ is the charge density of the plasma background. 
Since a persistent current (i.e., plasma background) cannot generate electromagnetic waves \citep{Yang18}, only the first two terms contribute to the radiation. Therefore, the radiation of the bunch-cavity pair is consistent with that of a separated electron/positron pair discussed by \citet{Yang20}.
Based on the charge density distribution, the pulse profile is given by $A(t)=A_0(t)-A_0(t-d/c)$, where $A_0(t)$ and $-A_0(t-d/c)$ correspond to the pulse profiles of the bunch and the cavity, respectively. According to the time-shifting property of the Fourier transform, one has $A(\omega)=A_0(\omega)-A_0(\omega)e^{i\omega d/c}$.
Using Eq.(\ref{CRspec}) and $P_A(\omega)=2|A(\omega)|^2$ by Eq.(\ref{PAomega}), one has
\citep[also see][]{Yang20}
\be
P_A(\omega)\propto4\sin^2\left(\frac{\omega}{\omega_d}\right)\omega^{2/3}e^{-\omega/\omega_c} ~~~\text{with }\omega_d\simeq 2c/d. \label{pairspectrum}
\ee
For $\omega\ll\omega_d$, one has $\sin^2(\omega/\omega_d)\propto\omega^2$, leading to $P_A(\omega)\propto\omega^{8/3}$ at the low-frequency band. Thus, the radiation spectrum is much narrower and harder than that of a point source given by Eq.(\ref{CRspec}). The emission spectrum is shown by the blue curve in Figure \ref{PA_spectra}.
Furthermore, it can be further proved that the above formula is also available for some more complex bunch-cavity systems. For example, \citet{Melikidze00} proposed that some plasma solitons with net charges will result from a ponderomotive Miller force. Each soliton consists of one large bunch and two small cavities (see Figure 2 in \citet{Melikidze00} and panel (d) of Figure \ref{bunch_shape}), because the excess of one charge is compensated by the lack of this charge in the nearby regions. The charge density distribution of the bunch-cavity system could be roughly described by $\rho_{\rm bc}(x)=-q\delta(x-d)+2q\delta(x)-q\delta(x+d)+\rho_0$. Similar to the calculation of the bunch-cavity pair, the same result as Eq.(\ref{pairspectrum}) is obtained, as shown by the blue curve in Figure \ref{PA_spectra}.

The above discussion assumes that the particles in the bunch have a Lorentz factor $\gamma$. If the energy distribution of the charged particles satisfies a power-law distribution, the corresponding radiation spectrum would be characterized by a multi-segment broken power-law, and the details have been discussed by \citet{Yang18}. 

\subsection{Radiation by a single fluctuating bunch}\label{subsection_fluctuating}

If the bunch lifetime is short, $\tau_B <\rho/\gamma c$, the observed pulse duration will be shorter than $\tau_c$ given by Eq.(\ref{frequency1}), leading to a higher typical frequency than that of classical curvature radiation, i.e.
\be
\tilde\omega_c\sim\left[\tau_B\left(1-\frac{v}{c}\right)\right]^{-1}\sim\frac{2\gamma^2}{\tau_B}.
\ee
A fluctuating bunch with a short lifetime will generate electromagnetic radiation with a higher frequency than the typical frequency of classical curvature radiation. 

We consider that a bunch forms and disperses intermittently when the building particles move along a field line, and the coherent radiation pulses are generated when the bunch exists, as shown in panel (b) of Figure \ref{bunch_fluctuation}. The bunch forms with a rate of $\lambda_B$ and disperses during a lifetime of $\tau_B$. 
Due to the relativistic motion of the building particles with $\gamma\gg1$, the pulse rate $\lambda_b$ and the pulse duration $\tau_b$ should be corrected by the propagation time-delay effect, i.e. 
\begin{align}
&\lambda_b=\left(1-\frac{v}{c}\right)^{-1}\lambda_B\simeq2\gamma^2\lambda_B,\\
&\tau_b=\left(1-\frac{v}{c}\right)\tau_B\simeq\frac{\tau_B}{2\gamma^2}\sim\tilde\omega_c^{-1},
\end{align}
where the factor of $(1-v/c)$ is due to the propagation time-delay effect.

\begin{figure}
    \centering
	\includegraphics[width = 1.0\linewidth, trim = 100 170 80 120, clip]{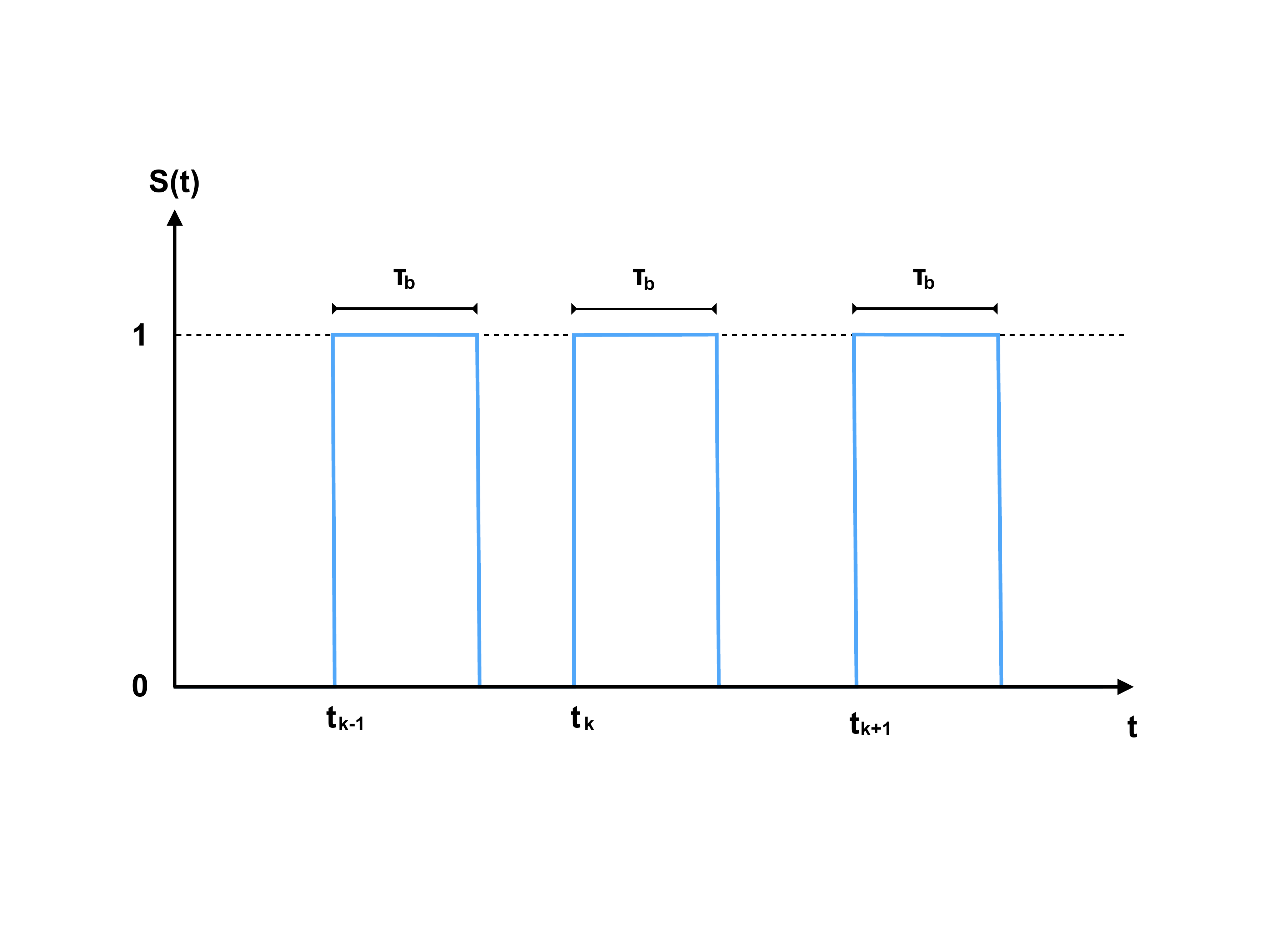}
    \caption{The pulse sampling function $S(t)$ given by Eq.(\ref{Sfunction}). $\{t_k\}$ is the starting time of the $k$-th pulse. Each pulse has a duration of $\tau_b$.}\label{Stplot} 
\end{figure}

During the time of $\rho/\gamma c$ when the emission cone sweeps the observing direction, the bunch would disperse and generate multiple times when the building plasma particles move along the magnetic field line. 
We consider that the radiation is in the form of $\tilde A(t)$. When the bunch exists, $\tilde A(t)\simeq A(t)$, where $A(t)$ is the radiation form of the classical curvature radiation by a single persistent source, as discussed in Section \ref{subsection_persistent}; when the bunch disappears, $\tilde A(t)\simeq0$. Therefore, the form of $\tilde A(t)$ can be written as
\be
\tilde A(t)=A(t)S(t),
\ee
where $S(t)$ is the pulse sampling function with
\begin{align}
S(t)&=
\left\{
\begin{aligned}
&1&&\text{for bunch existing},\\
&0&&\text{for bunch disappearing}, 
\end{aligned}
\right.\nonumber\\
&=\sum_ks(t-t_k),\label{Sfunction}
\end{align}
and
\begin{align}
s(t)=
\left\{
\begin{aligned}
&1&&\text{for } 0\leqslant t\leqslant\tau_b,\\
&0&&\text{for otherwise}. 
\end{aligned}
\right. \label{sfunction}
\end{align}
Here $t_k$ corresponds to the starting time of the $k$-th generation of the bunch, and $\tau_b$ is the pulse duration of the radiation from a bunch.
The pulse sampling function $S(t)$ is shown in Figure \ref{Stplot}.

We assume that pulse generation satisfies a Poisson process, thus, the probability to generate $k$ pulses during time $t$ is
\be
P_k(t)=\frac{(\lambda_b t)^k}{k!}e^{-\lambda_b t},
\ee
where $\lambda_b=2\gamma^2\lambda_B$ is the pulse rate that is corrected for the propagation time-delay effect. Here
$\{t_k\}$ in Eq.(\ref{Sfunction}) is distributed in a Poisson distribution with the parameter $\lambda_b$, and the probability density function of $\{t_k\}$ is related to the probability that the $k$-th point occurs in a short interval at the arbitrary time $t$ for short $\Delta t$,
\begin{align}
p_{t_k}(t)\Delta t&=P(t<t_k<t+\Delta t)\nonumber\\
&=P_{k-1}(t)P_1(\Delta t)=P_{k-1}(t)\lambda_b\Delta t.
\end{align}
Thus, one obtains the probability density function for $\{t_k\}$, i.e.
\be
p_{t_k}(t)=\lambda_b P_{k-1}(t).
\ee

Before calculating the emission spectrum $P_{\tilde A}(\omega)$ of $\tilde A(t)$, we are first interested in its autocorrelation function $R_{\tilde A}(\tau)$.
Since $S(t)$ corresponds to a random sampling process, $R_{\tilde A}(\tau)$ could be described by
\begin{align}
R_{\tilde A}(\tau)&=\mathcal{E}\left[\tilde A(t)*\tilde A^\dagger(-t)\right]=\mathcal{E}[(A(t)S(t))*(A(-t)S(-t))^\dagger]\nonumber\\
&=\mathcal{E}\left[\int_{-\infty}^{\infty} A(t+\tau)S(t+\tau)A^\dagger(t)S^\dagger(t)dt\right]\nonumber\\
&=\int_{-\infty}^{\infty} A(t+\tau)A^\dagger(t)\mathcal{E}\left[S(t+\tau)S^\dagger(t)\right]dt,\label{autocorr1}
\end{align}
where 
$\tilde A(t)*\tilde A^\dagger(-t)$ denotes the autocorrelation function of $\tilde A(t)$, and
$\mathcal{E}[X]$ denotes the expectation of the random variable $X$ involved by the random process. The symbol ``*'' denotes the convolution operator, and the superscript ``$^\dagger$'' denotes the conjugation operator. In the above calculation, the property of $\int_{-\infty}^{\infty} f(t)f^\dagger(t-\tau)dt=\int_{-\infty}^{\infty} f(t+\tau)f^\dagger(t)dt$ is used based on variable substitution $t-\tau\rightarrow t$.
For the Poisson sampling process, the autocorrelation of $R_S(\tau)$ satisfies \citep{Franks81}
\begin{align}
R_{S}(\tau)&=\mathcal{E}\left[S(t+\tau)S^\dagger(t)\right]=\sum_k\sum_j\mathcal{E}\left[s(t+\tau-t_k)s(t-t_j)\right]\nonumber\\
&=\sum_{k=j}\int_{-\infty}^{\infty} s(t+\tau-\xi)s(t-\xi)p_{t_k}(\xi)d\xi\nonumber\\
&+\sum_{k\neq j}\int_{-\infty}^{\infty} s(t+\tau-\eta)p_{t_k}(\eta)d\eta\int_{-\infty}^{\infty} s(t-\sigma)p_{t_k}(\sigma)d\sigma\nonumber\\
&=(\lambda_b q_s)^2+\lambda_b r_s(\tau),\label{Rs}
\end{align}
where 
\begin{align}
&q_s\equiv\int_{-\infty}^{\infty} s(t)dt,\label{qs}\\
&r_s(\tau)\equiv\int_{-\infty}^{\infty} s(t+\tau)s(t)dt.\label{rs}
\end{align}
Notice that both $S(t)$ and $s(t)$ are real functions according to Eq.(\ref{Sfunction}) and Eq.(\ref{sfunction}).
Since the autocorrelation function $R_{S}(\tau)$ is independent of $t$ (i.e., a wide-sense-stationary process), Eq.(\ref{autocorr1}) could be finally written as
\be
R_{\tilde A}(\tau)=R_{A}(\tau)R_{S}(\tau),\label{autocorrproduct}
\ee
where $R_A(\tau)$ is the autocorrelation function of $A(t)$. Therefore, the autocorrelation function of the product of $A(t)$ and $S(t)$ is the product of the autocorrelation of each one.

The emission spectrum of $\tilde A(t)$ is given by Eq.(\ref{PAomega}),
\be
P_{\tilde A}(\omega)\equiv\frac{d\tilde W}{d\Omega d\omega}
=2|\tilde A(\omega)|^2.\label{power1}
\ee
According to the convolution theorem and conjugation property of the Fourier transform, $|\tilde A(\omega)|^2$ could be written as 
\begin{align}
|\tilde A(\omega)|^2&=\tilde A(\omega)\tilde A^\dagger(\omega)\nonumber\\
&=\frac{1}{\sqrt{2\pi}}\mathcal{F}(\tilde A(t)*\tilde A^\dagger(-t))=\frac{1}{\sqrt{2\pi}}\mathcal{F}(R_{\tilde A}(\tau)),\label{WK}
\end{align}
where $\mathcal{F}(...)$ denotes the Fourier transform, the factor of $1/\sqrt{2\pi}$ is involved due to the definition of Fourier transform Eq.(\ref{Aomega}) and Eq.(\ref{At}). Thus, the Fourier transform of the autocorrelation function is the power spectrum, known as the Wiener-Khinchin's theorem. 
According to Eq.(\ref{autocorrproduct}), Eq.(\ref{power1}), Eq.(\ref{WK}) and the convolution theorem, the emission spectrum of $\tilde A(t)$ could be finally written as 
\begin{align}
P_{\tilde A}(\omega)&=\frac{2}{\sqrt{2\pi}}\mathcal{F}(R_{A}(\tau)R_{S}(\tau))=2|A(\omega)|^2*\frac{1}{\sqrt{2\pi}}\mathcal{F}(R_{S}(\tau))\nonumber\\
&=P_{ A}(\omega)*P_S(\omega),\label{VIP}
\end{align}
where $\mathcal{F}(R_{A}(\tau)R_{S}(\tau))=(1/\sqrt{2\pi})\mathcal{F}(R_{A}(\tau))*\mathcal{F}(R_{S}(\tau))$ is used, and the emission spectrum of the pulse sampling function $S(t)$ is defined as
\be
P_S(\omega)\equiv\frac{1}{\sqrt{2\pi}}\mathcal{F}(R_{S}(\tau)).\label{Ps}
\ee
Equation (\ref{VIP}) is the most important formula in this section, and we will use it to analyze the spectral features of the coherent radiation by fluctuating bunches. In the following discussion, we will discuss two mathematical models of the pulse sampling profile $S(t)$: impulsive sampling profile and rectangular sampling profile.

\subsubsection{Impulsive sampling profile}\label{subsubsection_impulsive}

If the pulse duration $\tau_b$ is much shorter than $\lambda_b^{-1}$, i.e., $\tau_b\lambda_b\ll1$, the pulse profile $s(t)$ in Eq.(\ref{sfunction}) can be well described using the delta function $\delta(t)$, $s(t)\simeq\tau_b\delta(t)$ for $\tau_b\rightarrow0$. According to Eq.(\ref{qs}) and Eq.(\ref{rs}), one has $r_s\simeq\tau_b^2\delta(\tau)$ and $q_s\simeq\tau_b$. Using Eq.(\ref{Sfunction}) and Eq.(\ref{Rs}), the autocorrelation function $R_{S}(\tau)$ of $S(t)$ is 
\be
R_S(\tau)=(\lambda_b\tau_b)^2+\lambda_b\tau_b^2\delta(\tau).
\ee
According to Eq.(\ref{Ps}), the emission spectrum of $S(t)$ is
\be
P_S(\omega)=(\lambda_b\tau_b)^2\delta(\omega)+\frac{\lambda_b\tau_b^2}{2\pi}.
\ee
Based on Eq.(\ref{VIP}), the emission spectrum of $\tilde A(t)$ is
\be
P_{\tilde A}(\omega)=P_A(\omega)*P_S(\omega)=(\lambda_b\tau_b)^2P_A(\omega)+\frac{\lambda_b\tau_b^2}{2\pi} P_{A,{\rm tot}},
\ee
where $P_{A,{\rm tot}}$ is the total radiation energy of $A(t)$,
\begin{align}
P_{A,{\rm tot}}=\int_{-\infty}^{\infty} P_A(\omega)d\omega.
\end{align}
Compared with that of a persistent bunch, $P_A(\omega)$, the emission spectrum of a fluctuating bunch is suppressed by a factor of $\sim(\lambda_b\tau_b)^2=(\lambda_B\tau_B)^2$. For the scenario of an impulsive sampling profile, $\lambda_b\tau_b\ll1$ has been potentially assumed.
Meanwhile, the emission spectrum of a single fluctuating bunch is the sum of the emission spectrum of a persistent bunch and a white noise that is independent of frequency $\omega$. 
The ``signal-to-noise ratio'' at the peak frequency $\omega_{\rm peak}$ in the frequency domain is given by
\be
\left.\frac{S}{N}\right|_{\rm peak}\simeq\frac{(\lambda_b\tau_b)^2P_A(\omega_{\rm peak})}{(\lambda_b\tau_b^2/2\pi)P_{A,{\rm tot}}}\simeq\frac{2\pi\lambda_b}{\omega_{\rm peak}},
\ee
where $P_{A,{\rm tot}}\sim \omega_{\rm peak}P_A(\omega_{\rm peak})$ is adopted based on the wide spectrum property of the curvature radiation. We can see that $(S/N)_{\rm peak}$ is independent of $\tau_b$, and the larger the pulse rate $\lambda_b$, the larger $(S/N)_{\rm peak}$. 
The spectrum $P_A(\omega)$ and its peak frequency $\omega_{\rm peak}$ depend on the spatial structure of the charged bunch as discussed in Section \ref{subsection_persistent}. For example: (1) If the bunch is a point-source, $P_A(\omega)$ is given by Eq.(\ref{CRspec}) and the peak frequency is $\omega_{\rm peak}\sim\omega_c\sim\gamma^3c/\rho$; (2) If the bunch is one-dimensional, $P_A(\omega)$ is given by Eq.(\ref{1Dbunch}) and the peak frequency is $\omega_{\rm peak}\sim\omega_l\sim2c/l$; (3) If the bunch is a bunch-cavity system, $P_A(\omega)$ is given by Eq.(\ref{pairspectrum}) and the peak frequency is $\omega_{\rm peak}\sim\omega_c$.

If one observes a non-white-noise signal in the frequency domain, i.e., $(S/N)_{\rm peak}\gg1$, $\lambda_b\gg\omega_{\rm peak}$ is required. For the GHz signal with $\omega_{\rm peak}/2\pi\sim10^9~{\rm rad~s^{-1}}$, the bunch formation rate should be $\lambda_b\gg 10^{9}~{\rm s}^{-1}$, leading to $\lambda_B\simeq\lambda_b/2\gamma^2\gtrsim 10^{3}~{\rm s^{-1}}~\gamma_3^{-2}$.
In particular, for an FRB with a typical duration of a few milliseconds, at least one bunch is produced during $\Delta t\sim1~{\rm ms}$, leading to $\lambda_B\gtrsim1/\Delta t\sim10^3~{\rm s^{-1}}$ and $\lambda_b\simeq2\gamma^2\lambda_B\gtrsim10^9~{\rm s^{-1}}\gamma_3^2\sim\omega_{\rm peak}$ and $(S/N)_{\rm peak}\gtrsim1$. Thus, the white-noise signal might not be significant for an FRB, if the FRB is produced by the bunch with a Lorentz factor $\gamma\gtrsim10^3$. Notice that the above conclusion potentially assumes that $\tau_b\lambda_b\ll1$ for the impulsive sampling profile.

\subsubsection{Rectangular sampling profile}\label{subsubsection_rectangular}

\begin{figure}
    \centering
	\includegraphics[width = 1.0\linewidth, trim = 0 0 0 0, clip]{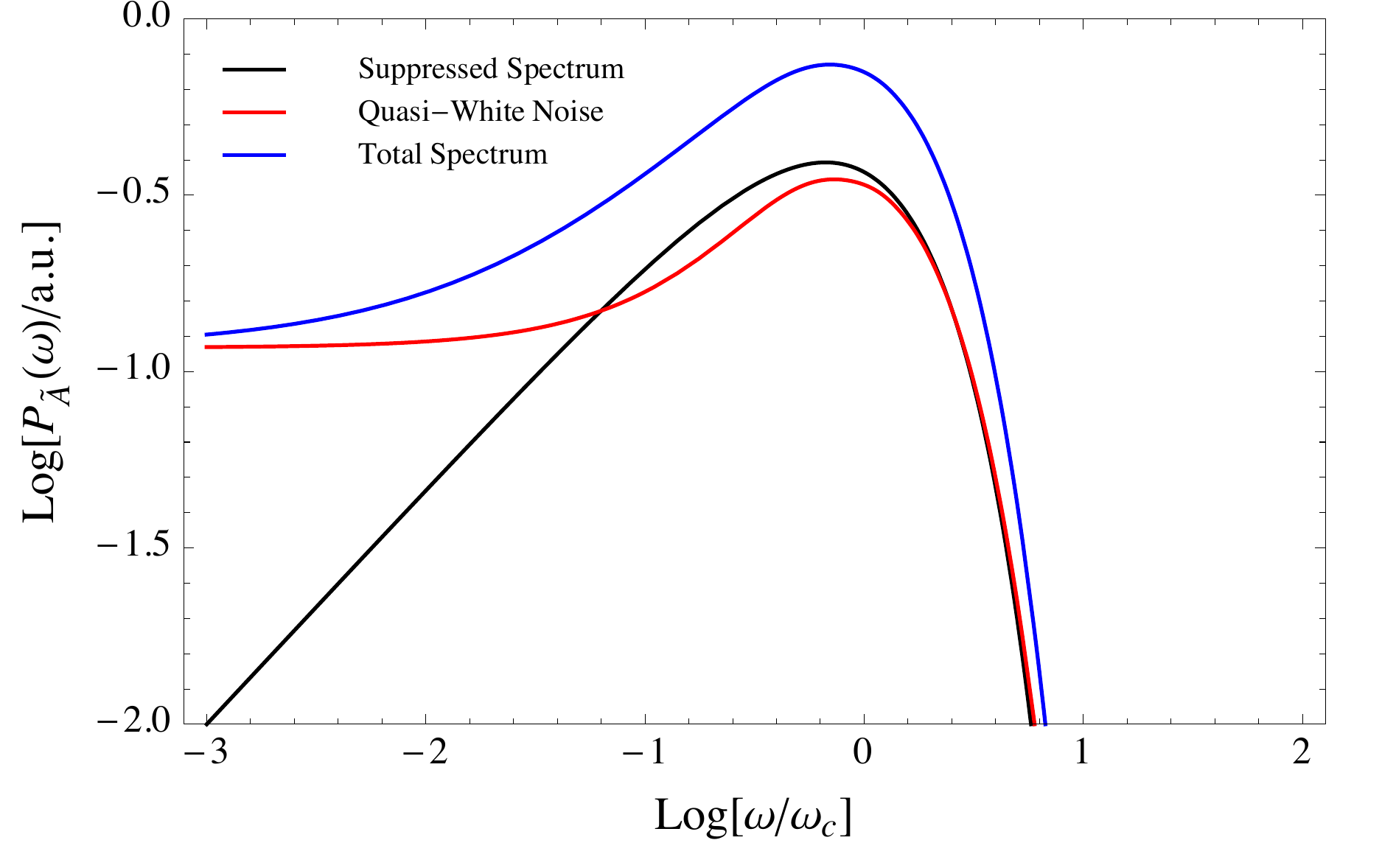}
	\includegraphics[width = 1.0\linewidth, trim = 0 0 0 0, clip]{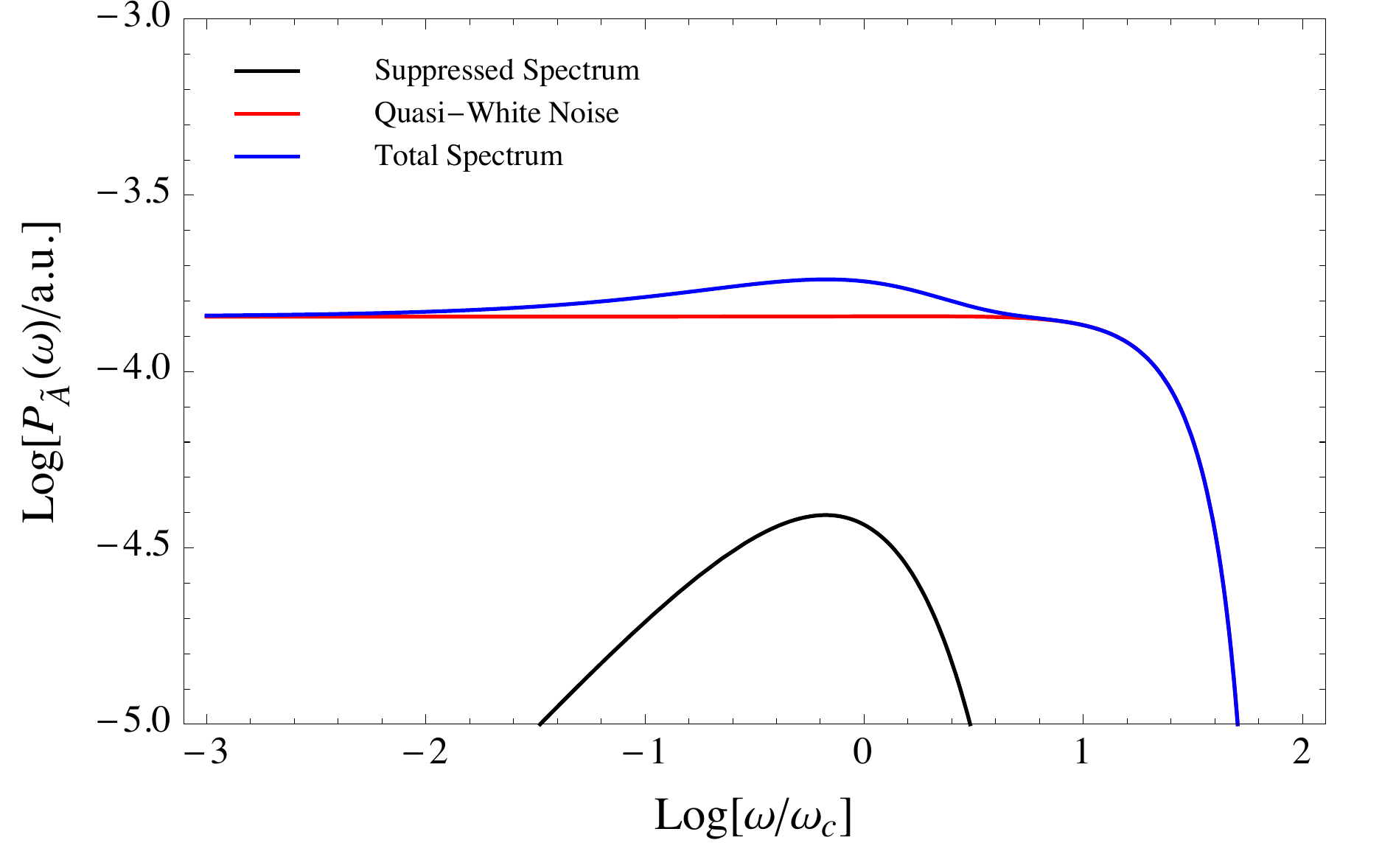}
    \caption{The emission spectrum of a single fluctuating bunch. The black, red, and blue curves correspond to the suppressed emission spectrum $(\lambda_b\tau_b)^2P_A(\omega)$, the quasi-white noise $(\lambda_b\tau_b^2/2\pi) P_{A}(\omega)*{\rm sinc}^2\left(\tau_b\omega/2\right)$, and the total emission spectrum, respectively. The top panel is the case with $\lambda_b=0.1\omega_{\rm peak}$ and $\tau_b=10\omega_{\rm peak}^{-1}$, and the bottom panel is the case with $\lambda_b=0.1\omega_{\rm peak}$ and $\tau_b=0.1\omega_{\rm peak}^{-1}$. Here we take $P_A(\omega)$ as the emission spectrum of a single persistent source given by Eq.(\ref{CRspec}) and the peak frequency is $\omega_{\rm peak}=\omega_c$. The unit of the emission spectrum is arbitrary.}\label{PAt_spectra} 
\end{figure}

Next, we generally consider that the function $s(t)$ given by Eq.(\ref{sfunction}) could be well described by a rectangular profile with width $\tau_b$, i.e.
\begin{align}
s(t)={\rm rect}\left(\frac{t}{\tau_b}\right)\equiv
\left\{
\begin{aligned}
&1&&\text{for }\left|\frac{t}{\tau_b}\right|\leqslant\frac{1}{2},\\
&0&&\text{for }\left|\frac{t}{\tau_b}\right|>\frac{1}{2},
\end{aligned}
\right.
\end{align}
where ${\rm rect}(x)$ is the rectangular function. Using Eq.(\ref{Sfunction}) and Eq.(\ref{Rs}), the autocorrelation function $R_{S}(\tau)$ is 
\be
R_S(\tau)=(\lambda_b\tau_b)^2+\lambda_b\tau_b\Lambda\left(\frac{\tau}{\tau_b}\right),
\ee
where $\Lambda(x)$ is the triangular function
\begin{align}
\Lambda(x)\equiv
\left\{
\begin{aligned}
&1-|x| &&\text{for }|x|\leqslant1,\\
&0 &&\text{for otherwise}. 
\end{aligned}
\right. 
\end{align}
According to Eq.(\ref{Ps}), the emission spectrum of $S(t)$ is
\be
P_S(\omega)=(\lambda_b\tau_b)^2\delta(\omega)+\frac{\lambda_b\tau_b^2}{2\pi}{\rm sinc}^2\left(\frac{\tau_b\omega}{2}\right).
\ee
Using Eq.(\ref{VIP}), the emission spectrum of $\tilde A(t)$ is
\be
P_{\tilde A}(\omega)=(\lambda_b\tau_b)^2P_A(\omega)+\frac{\lambda_b\tau_b^2}{2\pi} P_{A}(\omega)*{\rm sinc}^2\left(\frac{\tau_b\omega}{2}\right). \label{PAtilde}
\ee
Compared with that of a single persistent bunch, the emission spectrum of a fluctuating bunch is suppressed by a factor of $\sim(\lambda_b\tau_b)^2=(\lambda_B\tau_B)^2$.
The signal-to-noise ratio at the peak frequency in the frequency domain is given by
\be
\left.\frac{S}{N}\right|_{\rm peak}=\frac{2\pi\lambda_b P_A(\omega_{\rm peak})}{P_{A}(\omega)*{\rm sinc}^2(\tau_b\omega/2)}.
\ee 
According to the property of the convolution of two pulse profiles, we have the following conclusions:
(1) If $\tau_b>\omega_{\rm peak}^{-1}$, one has $(S/N)_{\rm peak}\sim\lambda_b\tau_b=\lambda_B\tau_B$, and the cutoff frequency of the whole spectrum is at $\sim\omega_{\rm peak}$, see the top panel of Figure \ref{PAt_spectra}. 
(2) If $\tau_b<\omega_{\rm peak}^{-1}$, one has $(S/N)_{\rm peak}\sim\lambda_b/\omega_{\rm peak}$, and there is a high-frequency cutoff in the white noise at $\tau_b^{-1}\sim \tilde\omega_c$, see the bottom panel of Figure \ref{PAt_spectra}. 
In particular, when $\tau_b\rightarrow0$, one has ${\rm sinc}^2(\tau_b\omega/2)\sim1$, so the above results become the case  of an impulsive sampling profile as discussed in Section \ref{subsubsection_impulsive}.
In summary, for both cases, the cutoff frequency is at 
\be
\omega_{\rm cut}\sim\max(\omega_{\rm peak},\tau_b^{-1}),
\ee
and the signal-to-noise ratio at the peak frequency in the frequency domain is
\be
\left.\frac{S}{N}\right|_{\rm peak}\sim\frac{\lambda_b}{\min(\omega_{\rm peak},\tau_b^{-1})}.
\ee

\subsection{Radiation by multiple fluctuating bunches along a field line}\label{subsection_multiple}

Next, we discuss the radiation by multiple fluctuating bunches along a field line, that is, there is a bunch train (with more than one bunch) along the field line. 
Each bunch emits a radiation pulse with the same shape but with random arrival times.
We consider that the radiation by the first fluctuating bunch in the bunch train is $\tilde A(t)$, then the radiation by multiple fluctuating bunches could be described as
\be
\hat A(t)=\sum_j^N \tilde A(t-t_j),
\ee
where $t_j$ is the arrival time of the pulse generated by the $j$-th bunch, and $N$ is the total number of bunches along a field line.
Similar to the generation process of a bunch that has been considered to be a Poisson process as discussed above,
the distribution of multiple fluctuating bunches along a field line may also satisfy the Poisson process, i.e., $\{t_j\}$ satisfies 
\begin{align}p_{t_j}(t)=\lambda_T P_{j-1}(t),\label{traindistribution}
\end{align} 
where $\lambda_T^{-1}$ corresponds to the average time separation between two adjacent bunches. 

\begin{figure}
    \centering
	\includegraphics[width = 1.0\linewidth, trim = 0 0 0 0, clip]{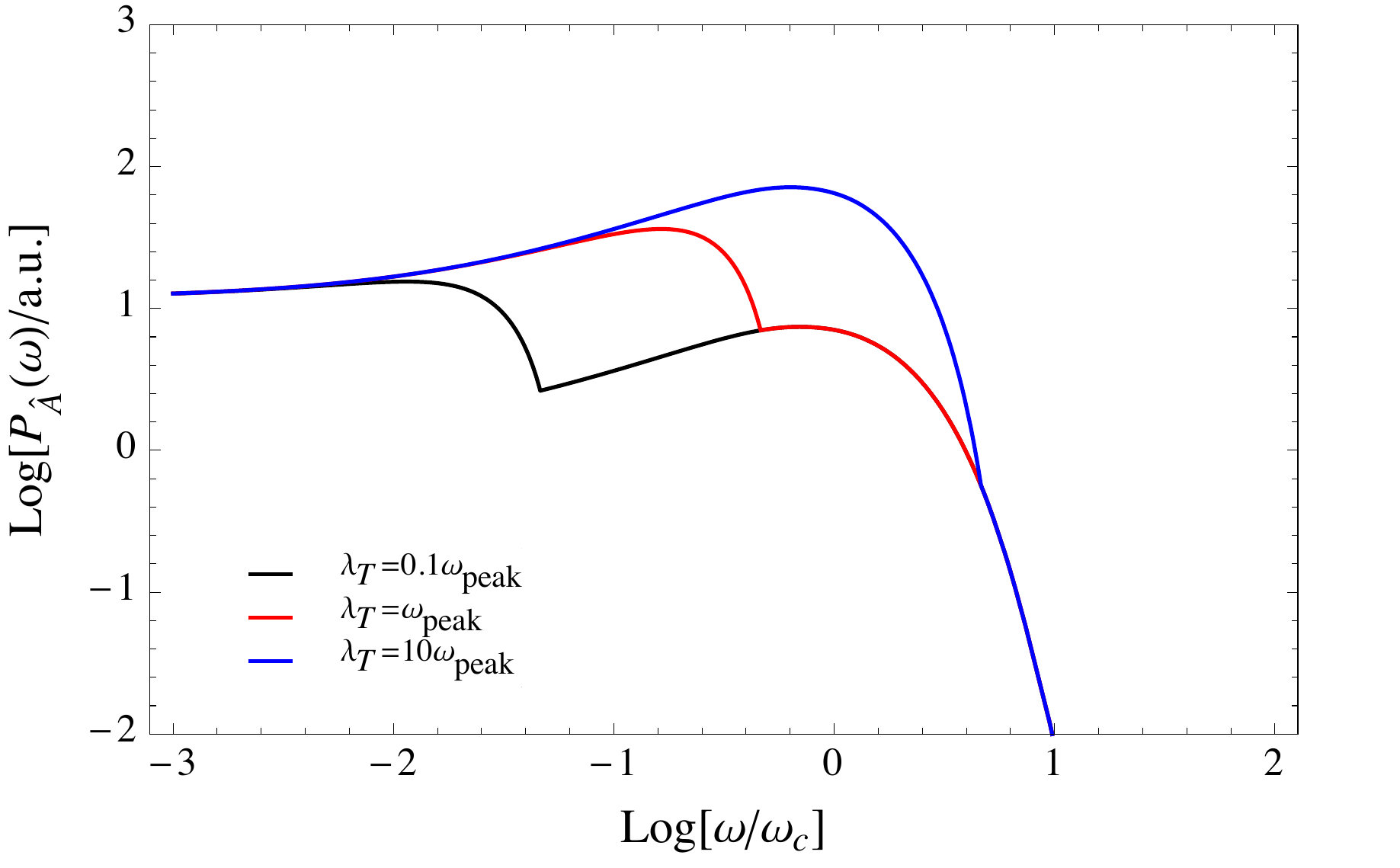}
    \caption{The emission spectrum of multiple fluctuating bunches. The black, red, and blue curves correspond to the cases with $\lambda_T=0.1\omega_{\rm peak}$, $\lambda_T=\omega_{\rm peak}$ and $\lambda_T=10\omega_{\rm peak}$, respectively. Here we take $P_{\tilde A}(\omega)$ as the emission spectrum of a fluctuating bunch given by Eq.(\ref{PAtilde}) (the corresponding emission spectrum $P_A(\omega)$ of the single persistent source is given by Eq.(\ref{CRspec})). The peak frequency is $\omega_{\rm peak}=\omega_c$.  $\lambda_b=0.1\omega_{\rm peak}$ and $\tau_b=10\omega_{\rm peak}^{-1}$ are taken. The unit of the emission spectrum is arbitrary. For clear display, we take the bunch number as $N=10$ in this figure. A larger value of $N$ would make the emission spectrum difference between the low-frequency coherent component and the high-frequency component more significant, and the break frequency is at $\sim2\lambda_T/\sqrt{N}$.}\label{multiple} 
\end{figure}

According to the time shifting property of Fourier transform, $\mathcal{F}[\tilde A(t-t_j)]=e^{i\omega t_j}\tilde A(\omega)$, the Fourier transform of $\hat A(t)$ is
\be
\hat A(\omega)=\mathcal{F}[\hat A(t)]=\tilde A(\omega)\sum_j^Ne^{i\omega t_j}.
\ee
Therefore, the emission spectrum of multiple fluctuating bunches is
\begin{align}
P_{\hat A}(\omega)=2|\hat A(\omega)|^2
=P_{\tilde A}(\omega)\left|\sum_j^Ne^{i\omega t_j}\right|^2.\label{coherence}
\end{align}
The coherence properties of the radiation by the bunch train are determined by the factor of $|\sum_j^Ne^{i\omega t_j}|^2$, and we will summarize its characteristics as follows:
(1) If the bunch separation is much larger than the wavelength, i.e., $\omega(t_j-t_{j-1})\sim\omega\lambda_T^{-1}\gg2\pi$, the phase factor $e^{i\omega t_j}$ would be randomly distributed in the complex number plane, leading to
\begin{align}
P_{\hat A}(\omega)=P_{\tilde A}(\omega)\left(N+\sum\sum_{j\neq k}e^{i\omega (t_j-t_k)}\right)\simeq NP_{\tilde A}(\omega),\label{PAhat1}
\end{align}
where $\sum\sum_{j\neq k}e^{i\omega (t_j-t_k)}\simeq 0$, because $\omega(t_j-t_k)$ are randomly distributed in $[0,2\pi]$. In this case, the radiation by multiple fluctuating bunches is the incoherent sum of those of many single fluctuating bunches, and the spectral shape of $P_{\hat A}(\omega)$ is the same as that of $P_{\tilde A}(\omega)$.
(2) If the bunch separation is much smaller than the wavelength, i.e., $\omega(t_j-t_{j-1})\sim\omega\lambda_T^{-1}\ll2\pi$, the sum in Eq.(\ref{coherence}) could be approximately calculated via the integration in the complex number plane, 
\begin{align}
\left|\sum_j^N e^{i\omega t_j}\right|^2
&\simeq\left|\lambda_{T}\int_0^{T} e^{i\omega t}dt\right|^2=N^2\left|\frac{e^{i\omega T}-1}{i\omega T}\right|^2\nonumber\\
&
=N^2\left[\frac{\sin(\omega/\omega_T)}{(\omega/\omega_T)}\right]^2~~~\text{with}~\omega_T=\frac{2}{T}=\frac{2\lambda_T}{N},\label{phasefactor}
\end{align}
where $T=N/\lambda_T$ is the bunch train duration (i.e., the duration time between the first bunch and the last bunch along a field line). One should note that the above equation gives $|\sum_j^Ne^{i\omega t_j}|^2=0$ when $\omega/\omega_T=n\pi$ with $n\in\mathbb{Z}^+$. The result of zero is due to the homogeneity and continuity assumptions in the integration calculation. The random process involved here will cause $|\sum_j^Ne^{i\omega t_j}|^2\sim N$ when $\omega/\omega_T=n\pi$. A Monte Carlo simulation with a distribution of Eq.(\ref{traindistribution}) could be used to test this result. Thus, the emission spectrum of multiple fluctuating bunches is
\begin{align}
P_{\hat A}(\omega)\simeq \max\left[N^2{\rm sinc}^2\left(\frac{\omega}{\omega_T}\right),N\right]P_{\tilde A}(\omega).\label{cohtrain}
\end{align}
According to Eq.(\ref{cohtrain}), the emission spectrum will become completely incoherent ($N$ term in the max function is dominated) for $\omega\gg2\lambda_T/\sqrt{N}$, meanwhile, Eq.(\ref{cohtrain}) is also applicable to Eq.(\ref{PAhat1}) in the case (1) due to $N\gg1$.
Furthermore, if the total bunch train length is much smaller than the wavelength, i.e., $\omega T\ll1$, one has $|\sum_j^Ne^{i\omega t_j}|^2\sim N^2$, leading to
\begin{align}
P_{\hat A}(\omega)\simeq N^2P_{\tilde A}(\omega).
\end{align} 
The emission from the bunch train is significantly coherent in this case. In Figure \ref{multiple}, we plot the radiation spectrum of the multiple fluctuating bunches with different values of $\lambda_T$ based on Eq.(\ref{cohtrain}). We can see the radiation is coherent for $\omega\lesssim2\lambda_T/\sqrt{N}$ and incoherent for $\omega\gtrsim2\lambda_T/\sqrt{N}$.

\section{Formation and dispersion of fluctuating bunches}\label{section_formation}

The above treatment has assumed continuous generation and dispersion of bunches. In this section, we discuss some mechanisms to produce bunches discussed in the literature.

In the magnetosphere of a neutron star, charged bunches could be generated by the development of a two-stream instability in a non-stationary plasma. For a pair plasma, there are three possible ways by which a two-stream instability can develop in the magnetosphere: (1) The two-stream instability is driven by the interaction between a high-energy beam and a secondary pair plasma \citep{Ruderman75,Benford77}. Such a mechanism is known to be inefficient; (2) The two-stream instability is due to the longitudinal drift of electrons and positrons in the secondary pair plasma \citep{Cheng77}; (3) The two-stream instability is produced by the overlapping secondary plasma clouds with different velocities due to intermittent discharges at the polar gap \citep{Usov87,Ursov88}.

The two-stream instability leads to the formation of  electrostatic Langmuir waves. When a Langmuir wave propagates in the magnetosphere, the oscillating electric field of the Langmuir wave forms a longitudinal concentration of charges, i.e., charged bunches.
However, the formation of charge bunches capable of explaining coherent radio emission has the following fundamental difficulty \citep{Lominadze86,Melikidze00,Lakoba18}. On one hand, the typical length $l$ of a bunch should be smaller than half of the wavelength $\lambda/2$ of the electromagnetic wave, $l<\lambda/2$, which makes the coherent radiation significant; otherwise, different regions in the bunch would emit independently and hence incoherently (see also Section \ref{subsection_persistent} for the effect by the bunch length). Meanwhile, if the bunches are caused by linear Langmuir waves with a wavelength $\lambda_L$, the bunch size is about half of the wavelength of the Langmuir wave, 
$l\sim\lambda_L/2$. Thus, the coherence condition is $\lambda>\lambda_L$, leading to 
\begin{align}
\omega<\omega_L,\label{condition1}
\end{align} 
where the approximate dispersion relations of $\omega\simeq kc$ and $\omega_L\simeq k_Lc$ are considered here. 

On the other hand, the timescale of the radiative process, $2\gamma^2\omega_{\rm cr}^{-1}$, where $\omega_{\rm cr}=\omega_c~{\rm or}~\tilde\omega_c$ is the typical frequency corresponding to the radiation process, must be shorter than the timescale $\tau_B$ over which the bunch exists. The bunch lifetime $\tau_B$ can be estimated as follows: the bunch lifetime in the plasma rest frame corresponds to the half of the period of the plasma oscillation, $\tau_B'\sim\pi\gamma^{1/2}\omega_p^{-1}$, where $\omega_p$ is the plasma frequency in the observer frame, and the factor of $\gamma^{1/2}$ is attributed to the density compression in the Lorentz transformation. In the observer frame, the bunch's lifetime is 
\be
\tau_B\sim\gamma\tau_B'\sim\pi\gamma^{3/2}\omega_p^{-1}\sim2\pi\gamma^2\omega_L^{-1},\label{LangtauB}
\ee 
where $\omega_L\sim2\gamma^{1/2}\omega_p$ is the Langmuir wave frequency\footnote{Note that the Langmuir wave frequency $\omega_L$ describes the wave oscillation frequency at a certain spatial position that does not directly reflect the bunch lifetime in the observer frame.}, see the following discussion.
Thus, this condition approximately requires $\omega_{\rm cr}\gtrsim\omega_L$. If the observed frequency is mainly at $\omega\sim\omega_{\rm cr}$, one finally has
\begin{align}
\omega\gtrsim\omega_L.\label{condition2}
\end{align}
The above two conditions of Eq.(\ref{condition1}) and Eq.(\ref{condition2}) are in contradiction with each other. Thus, the bunches created by the linear Langmuir wave cannot be a possible source of coherent curvature radiation. 

There might be some solutions to break the conditions of Eq.(\ref{condition1}) and Eq.(\ref{condition2}): (1) The charged bunches, such as ``charged solitons'' proposed by \citet{Melikidze00}, might exist for a duration much longer than the plasma oscillation period; (2) The bunch length is no longer related to the half-wavelength of Langmuir wave due to the non-linear effects; (3) The observed frequency $\omega$ is much smaller than the typical frequency $\omega_{\rm cr}$ that corresponds to the radiation process, $\omega\ll\omega_{\rm cr}$. In the last case, the observed frequency $\omega$ is no longer directly related to the radiative process, leading to both $\omega<\omega_L$ and $\omega_{\rm cr}\gtrsim\omega_L$ satisfied simultaneously (but there might be a possible absorption in the low-frequency band).

For the first two solutions, a non-linear Langmuir wave is preferred to produce the charged bunches contributing to significant coherent radio emission \citep[e.g.,][]{Melikidze00,Lakoba18,Rahaman22}.
The non-linear theory usually requires a priori a large amplitude for the Langmuir waves, i.e., the growth rate is sufficient for the breakdown of the linear theory \citep[e.g.,][]{Rahaman20}. 

We consider that the $\alpha$-th species in the plasma has the number density $n_\alpha$, charge $q_\alpha$, mass $m_\alpha$, dimensionless momenta $p_\alpha$, and  equilibrium distribution function $f_\alpha^{(0)}$.
The dispersion relation of the Langmuir wave of a one-dimensional relativistic plasma flow is given by \citep[e.g.,][]{Gedalin02,Asseo98,Rahaman20}
\be
1-\sum_\alpha\omega_{p,\alpha}^2\int \frac{1}{\gamma_\alpha^3}\frac{f_\alpha^{(0)}}{(\omega-\beta_\alpha kc)^2}dp_\alpha=0,\label{dispersion}
\ee
where $\omega_{p,\alpha}=(4\pi q_\alpha^2n_\alpha/m_\alpha)^{1/2}$ is the plasma frequency, and $\gamma_\alpha=(1+p_\alpha^2)^{1/2}$ is the Lorentz factor and $\beta_\alpha=p_\alpha/\gamma_\alpha$ is the dimensionless velocity. 
The cut-off frequency of the Langmuir waves corresponds to the long wavelength limit $k\rightarrow0$, leading to
\be
\omega_{\rm cut}^2=\sum_\alpha\omega_{p,\alpha}^2\int \frac{f_\alpha^{(0)}}{\gamma_\alpha^3}dp_\alpha,
\ee
and the characteristic frequency corresponds to the Langmuir mode touching the $\omega=kc$ line, leading to
\be
\omega_L^2=\sum_\alpha\omega_{p,\alpha}^2\int \frac{1}{\gamma_\alpha^3}\frac{f_\alpha^{(0)}}{(1-\beta_\alpha)^2}dp_\alpha.\label{chafreq}
\ee
We note that the Langmuir waves cannot exist for $\omega_{\rm cut}<\omega<\omega_L$ (superluminal waves), because the phase velocity of the Langmuir wave exceeds the light speed. If $\omega>\omega_L$ (subluminal waves), the phase velocity of the Langmuir wave is smaller than the light speed, which can cause the development of two-stream instability.
For example, if the distribution function of the pair plasma is close to a Gaussian distribution with a center of $\gamma_\pm$ and spread of $\Delta\gamma_\pm\ll\gamma_\pm$, i.e., the distribution function can be approximately described by the delta-function at $\gamma_\pm$, i.e., $f_\pm^{(0)}\sim\delta(p-p_\pm)$. The above two typical frequencies would become $\omega_{\rm cut}\sim\gamma_\pm^{-3/2}\omega_{p,\pm}$ and $\omega_{L}\sim2\gamma_\pm^{1/2}\omega_{p,\pm}$, where $\omega_\pm=(4\pi e^2n_\pm/m_e)^{1/2}$ is the pair plasma frequency, and $n_\pm$ is the number density of the pair plasma. 

Since the non-linear theory usually requires that the growth rate is sufficient for the breakdown of the linear theory, we first discuss the growth rate of two-stream instability by linear waves.
We consider that there are two plasma components (that are denoted by ``1'' and ``2'') with a relative motion along the magnetic field line,
Their typical Lorentz factors are $\gamma_1$ and $\gamma_2$ with $\gamma_1\gg\gamma_2$, and their number densities are $n_1$ and $n_2$, respectively. In the rest frame of each component, the plasma is assumed to be cold (the treatment for instability in a hot plasma could be seen in the recent work of \citet{Rahaman20,Melrose21}), leading to $f_j^{(0)}\simeq\delta(p-p_j)$ with $j=1,2$.
According to Eq.(\ref{dispersion}), the dispersion relation can be written as
\be
1-\frac{\omega_{p,1}^2}{\gamma_1^3(\omega-\beta_1 kc)^2}-\frac{\omega_{p,2}^2}{\gamma_2^3(\omega-\beta_2 kc)^2}=0,
\ee
where $\omega_{p,j}=(4\pi e n_j/m_e)^{1/2}$ is the plasma frequency of the component $j$.
We consider the resonant reactive instability as the bunching mechanism (the discussion for the non-resonant reactive instability and kinetic instability could be seen in \citet{Gedalin02,Rafat19b,Rahaman20}).
The solution of the above equation gives the growth rate \citep[e.g.,][]{Usov87,Gedalin02}
\be
\Gamma\sim\gamma_1^{-1}\gamma_2^{-1}\fraction{n_1}{n_2}{1/3}\omega_{L},\label{Gammarate}
\ee
and the characteristic frequency of the Langmuir wave is
\be
\omega_L\sim2\gamma_2^{1/2}\omega_{p,2},\label{omegalang}
\ee
In the following discussion, we analyze the bunch formation condition according to these results.

The amplitude of the linear Langmuir wave triggered by the two-stream instability depends on the gain 
\be
G=\Gamma\Delta t=\Gamma\frac{\Delta r}{c}\sim\Gamma\frac{r}{c},
\ee
where $\Delta t=\Delta r/c$ is the time available for the growth, and $\Delta r$ is the distance available for the growth along the field line.
The Langmuir wave amplitude is $\propto e^{G}$. 
In the magnetic field, the plasma density satisfies $n\propto r^{-m}$, where $m=3,4$ corresponds to the Goldreich-Julian magnetosphere \citep{Goldreich69} and the twisted magnetosphere, respectively (also see the following discussion). 
Thus, $n_1/n_2\sim\text{const}$ and $\omega_{p,2}\propto r^{-3/2}$ or $r^{-2}$ in Eq.(\ref{Gammarate}). The typical distance for the variation of the linear growth rate is $\Delta r\sim (2/3)r$ or $(1/2)r$ for $\Delta \omega_{p,2}\sim \omega_{p,2}$. We approximately take $G\sim\Gamma r/c$.

The non-linear theory requires a large amplitude of the Langmuir waves.
\citet{Rahaman20} proposed the following condition to indicate the breakdown of the linear theory: if the field energy density is equal to the total energy density, the linear theory would break down. Based on the energy distribution among Langmuir waves and particles in the plasma, \citet{Rahaman20} obtained a threshold gain indicating the breakdown of the linear regime,
\be
G_{\rm th}\simeq\frac{1}{2}\ln\sum_\alpha\gamma_\alpha^2.
\ee
Due to the logarithmic function in the above equation, the typical value of the gain threshold for the two-stream instability in various scenarios in the magnetosphere is about $G_{\rm th}\sim\text{a few}$.

The sufficient amplification would drive the system beyond the linear regime when $G= G_{\rm th}$, and the charged bunches contributing to significant coherent radio emission could be produced by the non-linear wave further. Thus, the bunch formation rate may be written as
\be
\lambda_B\sim\Gamma(G=G_{\rm th}),
\ee
and the bunch separation $l_T=c\lambda_T^{-1}$ corresponds to the wavelength of the Langmuir wave when $G\simeq G_{\rm th}$, i.e.
\be
l_T\equiv\frac{c}{\lambda_T}=\frac{2\pi c}{\omega_L}\sim\frac{\pi c}{\gamma_2^{1/2}\omega_{p,2}}
\ee
according to Eq.(\ref{omegalang}).
Due to the non-linear effect, the bunch size and lifetime would not be directly related to the Langmuir wave. 
For example, the charged solitons proposed by \citet{Melikidze00} could exist for a time much longer than the period of the Langmuir wave and the typical ripple size in one charged soliton also depends on the detailed properties of the soliton \citep{Melikidze00,Lakoba18,Rahaman22}.
Here, we briefly involve two parameters $(\zeta_l,\zeta_\tau)$ attributed to the non-linear effect to correct the bunch size $l$ and lifetime $\tau_B$,
\begin{align}
&l=\zeta_l\frac{2\pi c}{\omega_L}\sim\frac{\pi c\zeta_l}{\gamma_2^{1/2}\omega_{p,2}},\\
&\tau_B=\zeta_\tau\frac{2\pi\gamma^2}{\omega_L}\sim\frac{\gamma_2^{3/2}\pi \zeta_\tau}{\omega_{p,2}},
\end{align}
where the factor of $\gamma^2$ in $\tau_B$ is attributed to Eq. (\ref{LangtauB}).

Based on the above theory, as an example, we consider that the development of a two-stream instability is due to the non-stationarity of the plasma stream, and the pair plasma that flows out from the pulsar is inhomogeneous and gathers into separate clouds along the field lines \citep{Usov87,Ursov88,Asseo98}. 
The Lorentz factors of the electrons/positrons in the pair plasma have a distribution ranging from $\gamma_{\min}$ to $\gamma_{\max}$, so the pair clouds disperse as they flow out from the neutron star. The energy distribution of the plasma particles satisfies
\be
n(\gamma)d\gamma=n_\gamma\gamma^{-p}d\gamma\label{electrondistribution}
\ee
with $n_\gamma\equiv(p-1)\gamma_{\min}^{p-1}n_\pm$, where $n_\pm$ is the total pair number density.
At the interaction distance $r_i$, the high-energy particles with $\gamma\sim\gamma_{\max}$ of the cloud $B$ catch up with the low-energy particles with $\gamma\sim\gamma_{\min}$ of the cloud $A$ in front of the cloud $B$ with an initial separation $L$. The interaction distance is
\be
r_i=\frac{L}{v_{\max}-v_{\min}}\simeq2\gamma_{\min}^2L
\ee
for $\gamma_{\max}\gg\gamma_{\min}\gg1$, where $v_{\min}$ and $v_{\max}$ are the particle minimum and maximum velocities, respectively. Here we briefly assume that cloud $A$ and $B$ have the same number density.

At the distance $r_i$, there are approximately only particles with $\gamma\sim\gamma_{\min}$ and $\gamma\sim\gamma_{\max}$ in the merged cloud, so the two-stream instability could develop. 
Component 1 has a density $\sim\gamma_{\max}n(\gamma_{\max})$ and Lorentz factor $\gamma_{\max}$, and Component 2 has a density $\sim\gamma_{\min}n(\gamma_{\min})$ and Lorentz factor $\gamma_{\min}$. According to Eq.(\ref{Gammarate}) and Eq.(\ref{electrondistribution}), the growth rate of the two-stream instability is
\be
\Gamma\sim\gamma_{\min}^{\frac{2p-5}{6}}\gamma_{\max}^{-\frac{p+2}{3}}\omega_p,
\ee
where $\omega_p=(4\pi e n_e/m_e)^{1/2}$ and $n_e\sim \gamma_{\min}n(\gamma_{\min})$. 
In the magnetosphere, the pair plasma density is approximately given by
\be
n_\pm\sim\kappa\max(n_{\rm GJ},n_{\rm twist})\simeq\kappa\max\left(\frac{\Omega B(r)}{2\pi ec},\frac{B(r)\sin^2\theta\Delta\phi}{4\pi er}\right).
\ee
where $\kappa$ is the multiplicity of pair production, $n_{\rm GJ}=\Omega B(r)/2\pi ec$ is the Goldreich-Julian density \citep{Goldreich69}, $n_{\rm twist}=\nabla\times\vec{B}/4\pi e\sim B(r)\sin^2\theta\Delta\phi/4\pi er$ is the net charge number density in a twisted magnetosphere, $B(r)=B_p(r/R_n)^{-3}$ is the strength of a dipole field at the distance $r$, $B_p$ is the surface magnetic field, $R_n$ is the neutron star radius, and $\Omega$ is the neutron star angular velocity, $\theta$ is the poloidal angle and $\Delta\phi$ is the twisting angle of the field.

For example, we consider that a magnetar has a surface magnetic field $B_p\sim10^{14}~{\rm G}$, a rotation period $P\sim0.1~{\rm s}$, a twisting angle $\Delta\phi\sim0.1~{\rm rad}$, a polar angle $\sin^2\theta\sim0.1$, a multiplicity $\kappa\sim10^3$, and an outflowing plasma Lorentz factor $\gamma_\pm\sim100$. Because $n_{\rm twisting}\gg n_{\rm GJ}$ for the above typical parameters, the pair number density is $n_\pm\sim1.6\times10^{9}~{\rm cm^{-3}}r_8^{-4}$ and $n_e\sim\gamma_{\min}n(\gamma_{\min})\sim n_\pm$. 
We take $\gamma_{\min}\sim\gamma_\pm\sim100$, $\gamma_{\max}\sim10^4$ and $p\sim2.5$.
Thus, the linear growth rate for the development of the two-stream instability is $\Gamma\sim2.3\times10^{3}~{\rm s^{-1}}~r_8^{-2}$. 
Therefore, the non-linear Langmuir wave develops at $r\sim7.6\times10^8~{\rm cm}$ where $G\sim G_{\rm th}\sim O(1)$ is satisfied, and the correspond bunch formation rate is $\lambda_B\sim40~{\rm s^{-1}}$. The bunch separation is $l_T\sim\pi c/\gamma_\pm^{1/2}\omega_p\sim240~{\rm cm}$, and the bunch size and lifetime might be $l\sim24~{\rm cm}~\zeta_{l,-1}$ and $\tau_B\sim7.9\times10^{-3}~{\rm s}~\zeta_{\tau,2}$, respectively.  Charged bunches capable of contributing to coherent radio emission must be long-lived as proposed by \citet{Melikidze00,Rahaman22}.

\section{Conclusions and Discussions}\label{section_discussion}

Although coherent curvature radiation by charged bunches has been proposed to explain the coherent emissions of radio pulsars and FRBs, this mechanism still encounters some issues, including how the charged bunches form and disperse and what the radiation features are for the case of fluctuating bunches in the emission region.
In this work, we consider that the bunches in a neutron star magnetosphere form with an average rate of $\lambda_B$ and have an average lifetime of $\tau_B$.
We mainly analyze the spectral features of coherent curvature radiation by dynamically fluctuating bunches and discuss the possible physical mechanism for the formation and dispersion of the charged bunches in the magnetosphere of a neutron star. The following conclusions are drawn:

1. We first point out that the classical formula of calculating the brightness temperature of FRB emission, i.e. Eq.(\ref{brightness}), that involves the transient duration $\Delta t$ is not applicable to the scenario of the magnetospheric curvature radiation, because $\Delta t$ does not directly reflect the transverse size $l_e$ of the emission region for curvature radiation. Considering that the charged bunches move along the field line with a Lorentz factor $\gamma$ at a distance $r$ from the neutron star center, the transverse size of the emission region is estimated as $l_e\sim r/\gamma$ for $\omega\sim\omega_c$, leading to $l_e\ll c\Delta t$. Therefore, for the typical parameters of the magnetosphere, the brightness temperature should be much larger than that given by Eq.(\ref{brightness}).

2. The classical theory of  curvature radiation potentially assumes that the bunch lifetime satisfies $\tau_B>\rho/\gamma c$, where $\rho$ is the curvature radius and $\gamma$ is the bunch Lorentz factor. Both the typical frequency and the cutoff frequency are $\omega_c\sim\gamma^3c/\rho$ and the spectral feature depends on the spatial structure of the bunch. For example, we consider that the bunch has a lengthscale of $l$. Compared with the radiation spectrum of a single point-charge persistent bunch, the radiation spectrum by an extending bunch is  corrected by a factor of ${\rm sinc}^2(\omega/\omega_l)$ with $\omega_l\sim2c/l$, and the radiation power is suppressed by a factor of $(\omega_l/\omega_c)^{5/3}$. 
In particular, since the excess of one charge is usually compensated by the lack of this charge in the nearby regions, a bunch-cavity system might form in the magnetosphere. It has an emission spectrum much narrower than that of a single persistent bunch and with the emission spectrum of $P_A(\omega)\propto\omega^{8/3}$ in the low-frequency band.

3. If the bunch lifetime is short enough, $\tau_B <\rho/\gamma c$, the cutoff frequency would become $\tilde \omega_c\sim\gamma^2/\tau_B\gtrsim\omega_c$. Thus, a short-lived bunch will radiate electromagnetic waves with a higher frequency compared with that of classical curvature radiation. Considering that bunches form and disperse intermittently when the building plasma particles move along a magnetic field line, the emission spectrum of such a fluctuating bunch is a convolution between the emission spectrum of a single persistent bunch $P_{A}(\omega)$ and that of the pulse sampling function $P_S(\omega)$, $P_{\tilde A}(\omega)=P_A(\omega)*P_S(\omega)$, where $P_S(\omega)$ is the emission spectrum of the pulse sampling function $S(t)$ that is described by Eq. (\ref{Sfunction}) and Figure \ref{Stplot}.

4. According to the above point, we obtained the emission spectrum of a single fluctuating bunch, $P_{\tilde A}(\omega)$. 
We find that compared with that of a single fluctuating bunch, $P_{\tilde A}(\omega)$ is suppressed by a factor of $(\lambda_b\tau_b)^2$, where $\lambda_b\simeq2\gamma^2\lambda_B$ and $\tau_b\simeq\tau_B/2\gamma^2$ are the pulse rate and duration, respectively, $\lambda_B$ and $\tau_B$ are the bunch formation rate and lifetime, respectively, and the factor of $2\gamma^2$ is corrected by the propagation time-delay effect.
Meanwhile, there is a quasi-white noise in the wider band. 
We define $(S/N)_{\rm peak}$ as the ``signal-to-noise ratio'' at the peak frequency $\omega_{\rm peak}$ in the frequency domain, and $(S/N)_{\rm peak}\gtrsim1$ means that the spectrum is non-white-noise.
If $\tau_b<\omega_{\rm peak}^{-1}$, one has $(S/N)_{\rm peak}\sim\lambda_b/\omega_{\rm peak}$, and there is a high-frequency cutoff in the white noise at $\tau_b^{-1}\sim\tilde\omega_c$. 
If $\tau_b>\omega_{\rm peak}^{-1}$, one has $(S/N)_{\rm peak}\sim\lambda_b\tau_b=\lambda_B\tau_B$, the cutoff frequency of the whole spectrum is at $\sim \omega_{\rm peak}$. 

5. For multiple fluctuating bunches along a field line, if the bunch separation is longer than the wavelength, the emission spectrum of multiple fluctuating bunches would be the incoherent summation of that of each single fluctuating bunch. On the other hand, if the bunch separation is much smaller than the wavelength, the coherent radiation by multiple fluctuating bunches could be described by Eq.(\ref{cohtrain}), suggesting that the total radiation power is at least larger than the incoherent value.

At last, we also notice that the theory of the spectral analysis for fluctuating bunches not only applies to curvature radiation but also to coherent inverse Compton scattering (ICS) by charged bunches \citep{Zhang22}. 
This mechanism generally has a much higher radiation power than curvature radiation, so that a lower degree of coherence is needed to interpret FRBs.
Similar to the scenario of curvature radiation, a white-noise component in the emission spectrum would also appear due to bunch fluctuations. Compared with coherent curvature radiation, the major difference is that the emission spectrum of coherent ICS mainly depends on the properties of the incident electromagnetic waves which should be involved in the discussion of the emission of a single persistent bunch. On the other hand, since the radiation direction of coherent ICS is also along the magnetic field line due to the relativistic motion of the bunches, its emission spectrum might be modulated by the typical frequency $\omega_c$. A detailed analysis of this mechanism will be performed in the future. 

\section*{Acknowledgements}

We thank the anonymous referee for the helpful comments and suggestions.
We also thank Qiao-Chu Li for the constructive discussion about the signal theory and acknowledge helpful discussions with Ze-Nan Liu, Mordehai Milgrom and Yue Wu.
Y-PY's work is supported by the National Natural Science Foundation of China grant No.12003028, the National SKA Program of China (2022SKA0130100), and the China Manned Spaced Project (CMS-CSST-2021-B11). 

\section*{Data Availability}
This theoretical study did not generate any new data.

\bibliographystyle{mnras} 

\begin{thebibliography}{}
\makeatletter
\relax
\def\mn@urlcharsother{\let\do\@makeother \do\$\do\&\do\#\do\^\do\_\do\%\do\~}
\def\mn@doi{\begingroup\mn@urlcharsother \@ifnextchar [ {\mn@doi@}
  {\mn@doi@[]}}
\def\mn@doi@[#1]#2{\def\@tempa{#1}\ifx\@tempa\@empty \href
  {http://dx.doi.org/#2} {doi:#2}\else \href {http://dx.doi.org/#2} {#1}\fi
  \endgroup}
\def\mn@eprint#1#2{\mn@eprint@#1:#2::\@nil}
\def\mn@eprint@arXiv#1{\href {http://arxiv.org/abs/#1} {{\tt arXiv:#1}}}
\def\mn@eprint@dblp#1{\href {http://dblp.uni-trier.de/rec/bibtex/#1.xml}
  {dblp:#1}}
\def\mn@eprint@#1:#2:#3:#4\@nil{\def\@tempa {#1}\def\@tempb {#2}\def\@tempc
  {#3}\ifx \@tempc \@empty \let \@tempc \@tempb \let \@tempb \@tempa \fi \ifx
  \@tempb \@empty \def\@tempb {arXiv}\fi \@ifundefined
  {mn@eprint@\@tempb}{\@tempb:\@tempc}{\expandafter \expandafter \csname
  mn@eprint@\@tempb\endcsname \expandafter{\@tempc}}}

\bibitem[\protect\citeauthoryear{{Asseo} \& {Melikidze}}{{Asseo} \&
  {Melikidze}}{1998}]{Asseo98}
{Asseo} E.,  {Melikidze} G.~I.,  1998, \mn@doi [\mnras]
  {10.1046/j.1365-8711.1998.01990.x}, \href
  {https://ui.adsabs.harvard.edu/abs/1998MNRAS.301...59A} {301, 59}

\bibitem[\protect\citeauthoryear{{Bailes}}{{Bailes}}{2022}]{Bailes22}
{Bailes} M.,  2022, \mn@doi [Science] {10.1126/science.abj3043}, \href
  {https://ui.adsabs.harvard.edu/abs/2022Sci...378.3043B} {378, abj3043}

\bibitem[\protect\citeauthoryear{{Basu}, {Mitra}  \& {Melikidze}}{{Basu}
  et~al.}{2022}]{Basu22}
{Basu} R.,  {Mitra} D.,   {Melikidze} G.~I.,  2022, \mn@doi [\apj]
  {10.3847/1538-4357/ac5039}, \href
  {https://ui.adsabs.harvard.edu/abs/2022ApJ...927..208B} {927, 208}

\bibitem[\protect\citeauthoryear{{Beloborodov}}{{Beloborodov}}{2017}]{Beloborodov17}
{Beloborodov} A.~M.,  2017, \mn@doi [\apjl] {10.3847/2041-8213/aa78f3}, \href
  {https://ui.adsabs.harvard.edu/abs/2017ApJ...843L..26B} {843, L26}

\bibitem[\protect\citeauthoryear{{Benford} \& {Buschauer}}{{Benford} \&
  {Buschauer}}{1977}]{Benford77}
{Benford} G.,  {Buschauer} R.,  1977, \mn@doi [\mnras]
  {10.1093/mnras/179.2.189}, \href
  {https://ui.adsabs.harvard.edu/abs/1977MNRAS.179..189B} {179, 189}

\bibitem[\protect\citeauthoryear{{Buschauer} \& {Benford}}{{Buschauer} \&
  {Benford}}{1976}]{Buschauer76}
{Buschauer} R.,  {Benford} G.,  1976, \mn@doi [\mnras]
  {10.1093/mnras/177.1.109}, \href
  {https://ui.adsabs.harvard.edu/abs/1976MNRAS.177..109B} {177, 109}

\bibitem[\protect\citeauthoryear{{Cheng} \& {Ruderman}}{{Cheng} \&
  {Ruderman}}{1977}]{Cheng77}
{Cheng} A.~F.,  {Ruderman} M.~A.,  1977, \mn@doi [\apj] {10.1086/155105}, \href
  {https://ui.adsabs.harvard.edu/abs/1977ApJ...212..800C} {212, 800}

\bibitem[\protect\citeauthoryear{{Cooper} \& {Wijers}}{{Cooper} \&
  {Wijers}}{2021}]{Cooper21}
{Cooper} A.~J.,  {Wijers} R.~A.~M.~J.,  2021, \mn@doi [\mnras]
  {10.1093/mnrasl/slab099}, \href
  {https://ui.adsabs.harvard.edu/abs/2021MNRAS.508L..32C} {508, L32}

\bibitem[\protect\citeauthoryear{{Cordes} \& {Chatterjee}}{{Cordes} \&
  {Chatterjee}}{2019}]{Cordes19}
{Cordes} J.~M.,  {Chatterjee} S.,  2019, \mn@doi [\araa]
  {10.1146/annurev-astro-091918-104501}, \href
  {https://ui.adsabs.harvard.edu/abs/2019ARA&A..57..417C} {57, 417}

\bibitem[\protect\citeauthoryear{{Franks}}{{Franks}}{1981}]{Franks81}
{Franks} L.~E.,  1981, {Signal Theory (Revised Edition)}

\bibitem[\protect\citeauthoryear{{Gedalin}, {Gruman}  \& {Melrose}}{{Gedalin}
  et~al.}{2002}]{Gedalin02}
{Gedalin} M.,  {Gruman} E.,   {Melrose} D.~B.,  2002, \mn@doi [\mnras]
  {10.1046/j.1365-8711.2002.05922.x}, \href
  {https://ui.adsabs.harvard.edu/abs/2002MNRAS.337..422G} {337, 422}

\bibitem[\protect\citeauthoryear{{Gil}, {Lyubarsky}  \& {Melikidze}}{{Gil}
  et~al.}{2004}]{Gil04}
{Gil} J.,  {Lyubarsky} Y.,   {Melikidze} G.~I.,  2004, \mn@doi [\apj]
  {10.1086/379972}, \href
  {https://ui.adsabs.harvard.edu/abs/2004ApJ...600..872G} {600, 872}

\bibitem[\protect\citeauthoryear{{Ginzburg} \& {Zhelezniakov}}{{Ginzburg} \&
  {Zhelezniakov}}{1975}]{Ginzburg75}
{Ginzburg} V.~L.,  {Zhelezniakov} V.~V.,  1975, \mn@doi [\araa]
  {10.1146/annurev.aa.13.090175.002455}, \href
  {https://ui.adsabs.harvard.edu/abs/1975ARA&A..13..511G} {13, 511}

\bibitem[\protect\citeauthoryear{{Goldreich} \& {Julian}}{{Goldreich} \&
  {Julian}}{1969}]{Goldreich69}
{Goldreich} P.,  {Julian} W.~H.,  1969, \mn@doi [\apj] {10.1086/150119}, \href
  {https://ui.adsabs.harvard.edu/abs/1969ApJ...157..869G} {157, 869}

\bibitem[\protect\citeauthoryear{{Jackson}}{{Jackson}}{1998}]{Jackson98}
{Jackson} J.~D.,  1998, {Classical Electrodynamics, 3rd Edition}

\bibitem[\protect\citeauthoryear{{Katz}}{{Katz}}{2014}]{Katz14}
{Katz} J.~I.,  2014, \mn@doi [\prd] {10.1103/PhysRevD.89.103009}, \href
  {https://ui.adsabs.harvard.edu/abs/2014PhRvD..89j3009K} {89, 103009}

\bibitem[\protect\citeauthoryear{{Katz}}{{Katz}}{2018}]{Katz18}
{Katz} J.~I.,  2018, \mn@doi [\mnras] {10.1093/mnras/sty2459}, \href
  {https://ui.adsabs.harvard.edu/abs/2018MNRAS.481.2946K} {481, 2946}

\bibitem[\protect\citeauthoryear{{Kumar} \& {Bo{\v{s}}njak}}{{Kumar} \&
  {Bo{\v{s}}njak}}{2020}]{Kumar20}
{Kumar} P.,  {Bo{\v{s}}njak} {\v{Z}}.,  2020, \mn@doi [\mnras]
  {10.1093/mnras/staa774}, \href
  {https://ui.adsabs.harvard.edu/abs/2020MNRAS.494.2385K} {494, 2385}

\bibitem[\protect\citeauthoryear{{Kumar}, {Lu}  \& {Bhattacharya}}{{Kumar}
  et~al.}{2017}]{Kumar17}
{Kumar} P.,  {Lu} W.,   {Bhattacharya} M.,  2017, \mn@doi [\mnras]
  {10.1093/mnras/stx665}, \href
  {https://ui.adsabs.harvard.edu/abs/2017MNRAS.468.2726K} {468, 2726}

\bibitem[\protect\citeauthoryear{{Kumar}, {Gill}  \& {Lu}}{{Kumar}
  et~al.}{2022}]{Kumar22b}
{Kumar} P.,  {Gill} R.,   {Lu} W.,  2022, \mn@doi [\mnras]
  {10.1093/mnras/stac2446}, \href
  {https://ui.adsabs.harvard.edu/abs/2022MNRAS.516.2697K} {516, 2697}

\bibitem[\protect\citeauthoryear{{Lakoba}, {Mitra}  \& {Melikidze}}{{Lakoba}
  et~al.}{2018}]{Lakoba18}
{Lakoba} T.,  {Mitra} D.,   {Melikidze} G.,  2018, \mn@doi [\mnras]
  {10.1093/mnras/sty2152}, \href
  {https://ui.adsabs.harvard.edu/abs/2018MNRAS.480.4526L} {480, 4526}

\bibitem[\protect\citeauthoryear{{Liu}, {Wang}, {Yang}  \& {Dai}}{{Liu}
  et~al.}{2022}]{Liu22}
{Liu} Z.-N.,  {Wang} W.-Y.,  {Yang} Y.-P.,   {Dai} Z.-G.,  2022, \mn@doi [arXiv
  e-prints] {10.48550/arXiv.2212.13153}, \href
  {https://ui.adsabs.harvard.edu/abs/2022arXiv221213153L} {p. arXiv:2212.13153}

\bibitem[\protect\citeauthoryear{{Lominadze}, {Machabeli}, {Melikidze}  \&
  {Pataraya}}{{Lominadze} et~al.}{1986}]{Lominadze86}
{Lominadze} D.~C.,  {Machabeli} G.~Z.,  {Melikidze} G.~I.,   {Pataraya} A.~D.,
  1986, Soviet Journal of Plasma Physics, \href
  {https://ui.adsabs.harvard.edu/abs/1986SvJPP..12..712L} {12, 712}

\bibitem[\protect\citeauthoryear{{Longair}}{{Longair}}{2011}]{Longair11}
{Longair} M.~S.,  2011, {High Energy Astrophysics}

\bibitem[\protect\citeauthoryear{{Lu} \& {Kumar}}{{Lu} \& {Kumar}}{2018}]{Lu18}
{Lu} W.,  {Kumar} P.,  2018, \mn@doi [\mnras] {10.1093/mnras/sty716}, \href
  {https://ui.adsabs.harvard.edu/abs/2018MNRAS.477.2470L} {477, 2470}

\bibitem[\protect\citeauthoryear{{Lu}, {Kumar}  \& {Zhang}}{{Lu}
  et~al.}{2020}]{Lu20}
{Lu} W.,  {Kumar} P.,   {Zhang} B.,  2020, \mn@doi [\mnras]
  {10.1093/mnras/staa2450}, \href
  {https://ui.adsabs.harvard.edu/abs/2020MNRAS.498.1397L} {498, 1397}

\bibitem[\protect\citeauthoryear{{Luo}, {Zhu-Ge}  \& {Zhang}}{{Luo}
  et~al.}{2023}]{Luo23}
{Luo} J.-W.,  {Zhu-Ge} J.-M.,   {Zhang} B.,  2023, \mn@doi [\mnras]
  {10.1093/mnras/stac3206}, \href
  {https://ui.adsabs.harvard.edu/abs/2023MNRAS.518.1629L} {518, 1629}

\bibitem[\protect\citeauthoryear{{Lyubarsky}}{{Lyubarsky}}{2014}]{Lyubarsky14}
{Lyubarsky} Y.,  2014, \mn@doi [\mnras] {10.1093/mnrasl/slu046}, \href
  {https://ui.adsabs.harvard.edu/abs/2014MNRAS.442L...9L} {442, L9}

\bibitem[\protect\citeauthoryear{{Lyubarsky}}{{Lyubarsky}}{2021}]{Lyubarsky21}
{Lyubarsky} Y.,  2021, \mn@doi [Universe] {10.3390/universe7030056}, \href
  {https://ui.adsabs.harvard.edu/abs/2021Univ....7...56L} {7, 56}

\bibitem[\protect\citeauthoryear{{Lyubarsky} \& {Ostrovska}}{{Lyubarsky} \&
  {Ostrovska}}{2016}]{Lyubarsky16}
{Lyubarsky} Y.,  {Ostrovska} S.,  2016, \mn@doi [\apj]
  {10.3847/0004-637X/818/1/74}, \href
  {https://ui.adsabs.harvard.edu/abs/2016ApJ...818...74L} {818, 74}

\bibitem[\protect\citeauthoryear{{Melikidze}, {Gil}  \& {Pataraya}}{{Melikidze}
  et~al.}{2000}]{Melikidze00}
{Melikidze} G.~I.,  {Gil} J.~A.,   {Pataraya} A.~D.,  2000, \mn@doi [\apj]
  {10.1086/317220}, \href
  {https://ui.adsabs.harvard.edu/abs/2000ApJ...544.1081M} {544, 1081}

\bibitem[\protect\citeauthoryear{{Melrose}}{{Melrose}}{2017}]{Melrose17}
{Melrose} D.~B.,  2017, \mn@doi [Reviews of Modern Plasma Physics]
  {10.1007/s41614-017-0007-0}, \href
  {https://ui.adsabs.harvard.edu/abs/2017RvMPP...1....5M} {1, 5}

\bibitem[\protect\citeauthoryear{{Melrose}, {Rafat}  \& {Mastrano}}{{Melrose}
  et~al.}{2021}]{Melrose21}
{Melrose} D.~B.,  {Rafat} M.~Z.,   {Mastrano} A.,  2021, \mn@doi [\mnras]
  {10.1093/mnras/staa3324}, \href
  {https://ui.adsabs.harvard.edu/abs/2021MNRAS.500.4530M} {500, 4530}

\bibitem[\protect\citeauthoryear{{Metzger}, {Berger}  \& {Margalit}}{{Metzger}
  et~al.}{2017}]{Metzger17}
{Metzger} B.~D.,  {Berger} E.,   {Margalit} B.,  2017, \mn@doi [\apj]
  {10.3847/1538-4357/aa633d}, \href
  {https://ui.adsabs.harvard.edu/abs/2017ApJ...841...14M} {841, 14}

\bibitem[\protect\citeauthoryear{{Petroff}, {Hessels}  \& {Lorimer}}{{Petroff}
  et~al.}{2019}]{Petroff19}
{Petroff} E.,  {Hessels} J.~W.~T.,   {Lorimer} D.~R.,  2019, \mn@doi [\aapr]
  {10.1007/s00159-019-0116-6}, \href
  {https://ui.adsabs.harvard.edu/abs/2019A&ARv..27....4P} {27, 4}

\bibitem[\protect\citeauthoryear{{Qu}, {Zhang}  \& {Kumar}}{{Qu}
  et~al.}{2023}]{Qu23}
{Qu} Y.,  {Zhang} B.,   {Kumar} P.,  2023, \mn@doi [\mnras]
  {10.1093/mnras/stac3111}, \href
  {https://ui.adsabs.harvard.edu/abs/2023MNRAS.518...66Q} {518, 66}

\bibitem[\protect\citeauthoryear{{Rafat}, {Melrose}  \& {Mastrano}}{{Rafat}
  et~al.}{2019}]{Rafat19b}
{Rafat} M.~Z.,  {Melrose} D.~B.,   {Mastrano} A.,  2019, \mn@doi [Journal of
  Plasma Physics] {10.1017/S0022377819000643}, \href
  {https://ui.adsabs.harvard.edu/abs/2019JPlPh..85f9003R} {85, 905850603}

\bibitem[\protect\citeauthoryear{{Rahaman}, {Mitra}  \& {Melikidze}}{{Rahaman}
  et~al.}{2020}]{Rahaman20}
{Rahaman} S.~M.,  {Mitra} D.,   {Melikidze} G.~I.,  2020, \mn@doi [\mnras]
  {10.1093/mnras/staa2280}, \href
  {https://ui.adsabs.harvard.edu/abs/2020MNRAS.497.3953R} {497, 3953}

\bibitem[\protect\citeauthoryear{{Rahaman}, {Mitra}, {Melikidze}  \&
  {Lakoba}}{{Rahaman} et~al.}{2022}]{Rahaman22}
{Rahaman} S.~M.,  {Mitra} D.,  {Melikidze} G.~I.,   {Lakoba} T.,  2022, \mn@doi
  [\mnras] {10.1093/mnras/stac2264}, \href
  {https://ui.adsabs.harvard.edu/abs/2022MNRAS.516.3715R} {516, 3715}

\bibitem[\protect\citeauthoryear{{Ruderman}}{{Ruderman}}{1971}]{Ruderman71}
{Ruderman} M.,  1971, \mn@doi [\prl] {10.1103/PhysRevLett.27.1306}, \href
  {https://ui.adsabs.harvard.edu/abs/1971PhRvL..27.1306R} {27, 1306}

\bibitem[\protect\citeauthoryear{{Ruderman} \& {Sutherland}}{{Ruderman} \&
  {Sutherland}}{1975}]{Ruderman75}
{Ruderman} M.~A.,  {Sutherland} P.~G.,  1975, \mn@doi [\apj] {10.1086/153393},
  \href {https://ui.adsabs.harvard.edu/abs/1975ApJ...196...51R} {196, 51}

\bibitem[\protect\citeauthoryear{{Rybicki} \& {Lightman}}{{Rybicki} \&
  {Lightman}}{1986}]{Rybicki86}
{Rybicki} G.~B.,  {Lightman} A.~P.,  1986, {Radiative Processes in
  Astrophysics}

\bibitem[\protect\citeauthoryear{{Sturrock}}{{Sturrock}}{1971}]{Sturrock71}
{Sturrock} P.~A.,  1971, \mn@doi [\apj] {10.1086/150865}, \href
  {https://ui.adsabs.harvard.edu/abs/1971ApJ...164..529S} {164, 529}

\bibitem[\protect\citeauthoryear{{Tong} \& {Wang}}{{Tong} \&
  {Wang}}{2022}]{Tong22}
{Tong} H.,  {Wang} H.-G.,  2022, \mn@doi [Research in Astronomy and
  Astrophysics] {10.1088/1674-4527/ac71a5}, \href
  {https://ui.adsabs.harvard.edu/abs/2022RAA....22g5013T} {22, 075013}

\bibitem[\protect\citeauthoryear{{Ursov} \& {Usov}}{{Ursov} \&
  {Usov}}{1988}]{Ursov88}
{Ursov} V.~N.,  {Usov} V.~V.,  1988, \mn@doi [\apss] {10.1007/BF00638987},
  \href {https://ui.adsabs.harvard.edu/abs/1988Ap&SS.140..325U} {140, 325}

\bibitem[\protect\citeauthoryear{{Usov}}{{Usov}}{1987}]{Usov87}
{Usov} V.~V.,  1987, \mn@doi [\apj] {10.1086/165546}, \href
  {https://ui.adsabs.harvard.edu/abs/1987ApJ...320..333U} {320, 333}

\bibitem[\protect\citeauthoryear{{Wang}, {Jiang}, {Lee}, {Xu}  \&
  {Zhang}}{{Wang} et~al.}{2022a}]{Wang22c}
{Wang} W.-Y.,  {Jiang} J.-C.,  {Lee} K.,  {Xu} R.,   {Zhang} B.,  2022a,
  \mn@doi [\mnras] {10.1093/mnras/stac3070}, \href
  {https://ui.adsabs.harvard.edu/abs/2022MNRAS.517.5080W} {517, 5080}

\bibitem[\protect\citeauthoryear{{Wang}, {Yang}, {Niu}, {Xu}  \&
  {Zhang}}{{Wang} et~al.}{2022b}]{Wang22b}
{Wang} W.-Y.,  {Yang} Y.-P.,  {Niu} C.-H.,  {Xu} R.,   {Zhang} B.,  2022b,
  \mn@doi [\apj] {10.3847/1538-4357/ac4097}, \href
  {https://ui.adsabs.harvard.edu/abs/2022ApJ...927..105W} {927, 105}

\bibitem[\protect\citeauthoryear{{Xiao} \& {Dai}}{{Xiao} \&
  {Dai}}{2022}]{Xiao22}
{Xiao} D.,  {Dai} Z.-G.,  2022, \mn@doi [\aap] {10.1051/0004-6361/202142268},
  \href {https://ui.adsabs.harvard.edu/abs/2022A&A...657L...7X} {657, L7}

\bibitem[\protect\citeauthoryear{{Xiao}, {Wang}  \& {Dai}}{{Xiao}
  et~al.}{2021}]{Xiao21}
{Xiao} D.,  {Wang} F.,   {Dai} Z.,  2021, \mn@doi [Science China Physics,
  Mechanics, and Astronomy] {10.1007/s11433-020-1661-7}, \href
  {https://ui.adsabs.harvard.edu/abs/2021SCPMA..6449501X} {64, 249501}

\bibitem[\protect\citeauthoryear{{Yang} \& {Zhang}}{{Yang} \&
  {Zhang}}{2018a}]{Yang18b}
{Yang} Y.-P.,  {Zhang} B.,  2018a, \mn@doi [\apjl] {10.3847/2041-8213/aada4f},
  \href {https://ui.adsabs.harvard.edu/abs/2018ApJ...864L..16Y} {864, L16}

\bibitem[\protect\citeauthoryear{{Yang} \& {Zhang}}{{Yang} \&
  {Zhang}}{2018b}]{Yang18}
{Yang} Y.-P.,  {Zhang} B.,  2018b, \mn@doi [\apj] {10.3847/1538-4357/aae685},
  \href {https://ui.adsabs.harvard.edu/abs/2018ApJ...868...31Y} {868, 31}

\bibitem[\protect\citeauthoryear{{Yang} \& {Zhang}}{{Yang} \&
  {Zhang}}{2020}]{Yang20b}
{Yang} Y.-P.,  {Zhang} B.,  2020, \mn@doi [\apjl] {10.3847/2041-8213/ab7ccf},
  \href {https://ui.adsabs.harvard.edu/abs/2020ApJ...892L..10Y} {892, L10}

\bibitem[\protect\citeauthoryear{{Yang}, {Zhu}, {Zhang}  \& {Wu}}{{Yang}
  et~al.}{2020}]{Yang20}
{Yang} Y.-P.,  {Zhu} J.-P.,  {Zhang} B.,   {Wu} X.-F.,  2020, \mn@doi [\apjl]
  {10.3847/2041-8213/abb535}, \href
  {https://ui.adsabs.harvard.edu/abs/2020ApJ...901L..13Y} {901, L13}

\bibitem[\protect\citeauthoryear{{Zhang}}{{Zhang}}{2020}]{Zhang20}
{Zhang} B.,  2020, \mn@doi [\nat] {10.1038/s41586-020-2828-1}, \href
  {https://ui.adsabs.harvard.edu/abs/2020Natur.587...45Z} {587, 45}

\bibitem[\protect\citeauthoryear{{Zhang}}{{Zhang}}{2022a}]{Zhang22b}
{Zhang} B.,  2022a, \mn@doi [arXiv e-prints] {10.48550/arXiv.2212.03972}, \href
  {https://ui.adsabs.harvard.edu/abs/2022arXiv221203972Z} {p. arXiv:2212.03972}

\bibitem[\protect\citeauthoryear{{Zhang}}{{Zhang}}{2022b}]{Zhang22}
{Zhang} B.,  2022b, \mn@doi [\apj] {10.3847/1538-4357/ac3979}, \href
  {https://ui.adsabs.harvard.edu/abs/2022ApJ...925...53Z} {925, 53}

\bibitem[\protect\citeauthoryear{{Zhu-Ge}, {Luo}  \& {Zhang}}{{Zhu-Ge}
  et~al.}{2023}]{Zhu-Ge23}
{Zhu-Ge} J.-M.,  {Luo} J.-W.,   {Zhang} B.,  2023, \mn@doi [\mnras]
  {10.1093/mnras/stac3599}, \href
  {https://ui.adsabs.harvard.edu/abs/2023MNRAS.519.1823Z} {519, 1823}

\makeatother
\end{thebibliography}

\bsp	
\label{lastpage}

\end{document}